\documentclass[aps,pra,english,twocolumn,letterpaper,superscriptaddress,notitlepage]{revtex4-1}

\usepackage[T1]{fontenc}
\usepackage{graphicx}
\usepackage{epstopdf}
\usepackage{amssymb}
\usepackage{amsmath}
\usepackage[urlcolor=blue,hyperindex,colorlinks,bookmarks=true,linkcolor=red,citecolor=black]{hyperref}
\usepackage[normalem]{ulem}
\usepackage[abs]{overpic}
\usepackage{color}
\usepackage{rotating}
\usepackage{physics}
\usepackage{bm}
\usepackage{soul}

\usepackage[english]{babel}
\usepackage{blindtext,tikz}
\usetikzlibrary{calc}

\newtheorem{lemma}{Lemma}

\def\iden{\hat{\mathbb{I}}}

\newcommand{\qp}{\mathcal{E}}

\begin{document}

\title{Clifford group restricted eavesdroppers in quantum key distribution}

\author{L.~C.~G.~Govia}
\email{luke.c.govia@raytheon.com}
\affiliation{Raytheon BBN Technologies, 10 Moulton St., Cambridge, MA 02138, USA}
\author{D.~Bunandar}
\affiliation{Research Laboratory of Electronics, Massachusetts Institute of Technology, Cambridge, MA 02139, USA}
\author{J.~Lin}
\affiliation{Institute for Quantum Computing and Department of Physics and Astronomy, University of Waterloo, Waterloo, Ontario, Canada N2L 3G1}
\author{D.~Englund}
\affiliation{Research Laboratory of Electronics, Massachusetts Institute of Technology, Cambridge, MA 02139, USA}
\author{N.~L\"utkenhaus}
\affiliation{Institute for Quantum Computing and Department of Physics and Astronomy, University of Waterloo, Waterloo, Ontario, Canada N2L 3G1}
\author{H.~Krovi}
\affiliation{Raytheon BBN Technologies, 10 Moulton St., Cambridge, MA 02138, USA}

\begin{abstract}
  Quantum key distribution (QKD) promises provably secure cryptography, even to attacks from an all-powerful adversary. However, with quantum computing development lagging behind QKD, the assumption that there exists an adversary equipped with a universal fault-tolerant quantum computer is unrealistic for at least the near future. Here, we explore the effect of restricting the eavesdropper's computational capabilities on the security of QKD, and find that improved secret key rates are possible. Specifically, we show that for a large class of discrete variable protocols higher key rates are possible if the eavesdropper is restricted to a unitary operation from the Clifford group. Further, we consider Clifford-random channels consisting of mixtures of Clifford gates. We numerically calculate a secret key rate lower bound for BB84 with this restriction, and show that in contrast to the case of a single restricted unitary attack, the mixture of Clifford based unitary attacks does not result in an improved key rate.
\end{abstract}

\maketitle

\section{Introduction}
Quantum key distribution (QKD) offers a potential path to quantum-safe cryptography that does not rely on the conjectured hardness of computational problems. An important problem in QKD is to devise a protocol that can be implemented using existing technology while maximizing the achievable secret key generation rate. QKD analyses have, at most times, assumed that the eavesdropper is all-powerful: able to perform any quantum operation on the signals transferred between Alice and Bob. Such operations can, in principle, be implemented using a universal quantum computer along with a long term quantum memory. This security without assumptions on the computational power of the adversary is one of the most compelling reasons to replace classical key exchange protocols with QKD.

Large scale fault-tolerant quantum computers that can implement Shor's factoring algorithm \cite{Shor:1994aa} will render currently used classical cryptosystems such as the Rivest-Shamir-Adleman (RSA) protocol \cite{Rivest:1978aa}, elliptic curve cryptography \cite{Koblitz1987}, and Diffie-Hellmann key exchange \cite{Diffie:1976aa} insecure. Several classical cryptosystems, such as code-based encryption~\cite{McEliece1978}, lattice-based encryption~\cite{Hoffstein:1998}, and supersingular isogenies \cite{De-Feo:2014aa} have been proposed as replacements for RSA, and these post-quantum cryptosystems are based on hard problems that are conjectured to be difficult for quantum computers to solve efficiently~\cite{Bernstein:2017aa}. As with all classical cyptosystems, proof of their security relies on the computational hardness of the underlying mathematical problem, or equivalently, on the computational power limitations of the adversary. In fact, there is so far no proof that guarantees the nonexistence of an efficient quantum---or even classical---algorithm to break these post-quantum cryptosystems. Thus, in many real world scenarios, the computational power of the adversary is the physically relevant condition.

While promising security without computational assumptions, in practice QKD protocols have been very challenging to implement, with successful demonstrations within a range of only a few hundred kilometers in optical fibre, see e.g.~\cite{Korzh:2015aa,Yin:2016aa,Bunandar:2018aa}, and the rate without intermediate stations inevitably falls off with the transmittance of the channel \cite{Pirandola:2017cj}. Quantum repeaters \cite{Muralidharan:2016ye,Krovi2016, PhysRevA.92.022357} have been proposed as a way to improve over direct transmission \cite{PLOB, Takeoka2014}. While impressive progress has been made towards their demonstration, no such repeater has been realized \cite{RevModPhys.83.33}. However, assuming nothing about the nature of the adversary is overly pessimistic in the era of noisy intermediate-scale quantum (NISQ) computers~\cite{Preskill:2018}. Moreover, in the classical case, an eavesdropper can store the public information indefinitely, and attack in the future when better algorithms or computational devices become available. In the quantum case, without a long-lived quantum memory, the eavesdropper must attack during the key exchange. Thus, for near-term QKD implementations the more relevant concern is the computational power of the eavesdropper \emph{today}.

In this work, we examine the role that computational assumptions on the eavesdropper have on the secret key rate of discrete variable quantum key distribution (DV-QKD) protocols. We examine the secret key rates that can be achieved if one restricts the eavesdropper to have access to a subclass of quantum operations formed by the Clifford group. Clifford operations are not universal and circuits built from them can be efficiently simulated classically \cite{Gottesman:1999aa}, such that we would expect such a restriction to limit the power of the eavesdropper. We show that this is indeed the case for a large family of DV-QKD protocols. We then extend to Clifford-random channels, allowing the eavesdropper to implement a quantum channel described by a convex combination of unitary Clifford gates, and numerically calculate a lower bound to the secret key rate possible under this restriction for BB84 \cite{Bennett:1984aa}.

The problem of numerically computing the secret key rate can be cast as a nonlinear semi-definite program (SDP)~\cite{Coles:2016aa,Winick:2018aa}. In this framework, it becomes clear that the optimal attack an eavesdropper can use dictates the secret key rate that can be achieved. In particular, restricting the computational capabilities of the eavesdropper is equivalent to introducing further constraints to the key rate SDP. To our knowledge, such computational assumptions on the eavesdropper and their impact on the secret key rate have not been considered in the literature.

Some work that is similar in spirit considers restrictions to the eavesdropper's ability to perform a coherent attack, either due to a noisy quantum memory used for storage \cite{Damgaard:fq}, or general decoherence \cite{Hosseinidehaj:2019aa}. For these models, it has been shown that limiting the capabilities of the eavesdropper improves the key rate. Further, improved key rates can also be found when Eve's access to a wiretap channel is restricted, such as in a free-space communication setting where she can only collect a fraction of the transmitted optical signals \cite{Pan:2019aa}.

\section{Results}

\subsection{Clifford Group Restriction}
\label{sec:MUBs}

We consider the broad class of prepare-and-measure QKD protocols from Ref.~\cite{Ferenczi:2012aa}, for which the set of possible signal states consists of all elements from either 2, $d$, or $d+1$ mutually unbiased bases (MUBs), where $d$ (assumed prime) is the Hilbert space dimension of the signal-state system. Examples of QKD protocols contained in this class include BB84 \cite{Bennett:1984aa}, and the 6-state protocol \cite{Bruss:1998aa}. For these protocols, we assume that the eavesdropper Eve only has access to Bob's eventual half of the system, and use a source-replacement scheme (see for e.g.~\cite{Bennett:1992aa,Curty:2004aa,Ferenczi:2012aa}) to describe prepare-and-measure protocols as entanglement-based. We are interested in Eve's optimal attack that minimizes the secret key rate.

The key observation of this section is that for the protocols described in Ref.~\cite{Ferenczi:2012aa}, the state $\rho_{AB}$ describing the optimal attack is also the Choi state describing the effective channel Eve implements on Bob's system. We can then ask what computational resources are required to simulate this effective channel, and determine if a restricted Eve can implement the optimal attack. We show that a Clifford-restricted Eve cannot implement the optimal attack in all situations. In the following we present our results, with details of the derivation contained in the appendix \ref{app:CG}.

The optimal attack for the protocols described in Ref.~\cite{Ferenczi:2012aa} leaves the final state of Alice and Bob in what is known as a \emph{Bell-diagonal} state
\begin{align}
  \rho_{AB} = \sum_{r,s}^{d-1}b_{r,s}\ketbra{B^d_{r,s}},
\end{align}
where $b_{r,s}$ are related to the bit-error rate and satisfy $\sum_{r,s}^{d-1}b_{r,s} = 1$ and $0 \leq b_{r,s} \leq 1$. Here $\ket{B^d_{r,s}}$ (see appendix \ref{app:CG}) is a generalized Bell state. We now show how we can recast this expression in the form
\begin{align}
  \rho_{AB} = \rho_{\qp_{B}} = \mathcal{I}\otimes\qp_B\left(\ketbra{B^d_{0,0}}\right), \label{eqn:Choi}
\end{align}
where $\rho_{\qp_{B}}$ is the Choi dual-state describing the effective process on Bob's qubit due to Eve's interaction.

We recall that the Choi dual-state of a quantum process $\qp$ is defined by the action of the process on half of the maximally entangled state $\ket{B^d_{0,0}}$ \cite{Nielsen00}. Using that
\begin{align}
  \ket{B^d_{r,s}} = \iden_d\otimes\hat{P}^d_{r,s}\ket{B^d_{0,0}},
\end{align}
where $\hat{P}^d_{r,s}$ are generalized Pauli matrices (see appendix \ref{app:CG}), we rewrite the optimal attack as
\begin{align}
  \rho_{AB} = \sum_{r,s}^{d-1}b_{r,s}~\iden_d\otimes\hat{P}^d_{r,s}\ketbra{B^d_{0,0}}\iden_d\otimes\left(\hat{P}^d_{r,s}\right)^\dagger, \label{eqn:local}
\end{align}
which has the desired form of a Choi dual-state, demonstrating that $\rho_{AB} = \rho_{\qp_{B}}$.

From here, it is straightforward to see that the effective channel $\qp_B$ on Bob's system resulting from Eve's optimal attack is
\begin{align}
  \qp_B(\rho_B) = \sum_{r,s}^{d-1}b_{r,s}\hat{P}^d_{r,s}\rho_B\left(\hat{P}^d_{r,s}\right)^\dagger. \label{eqn:channelB}
\end{align}
We refer to this as a \emph{Pauli-random} channel, as it is the convex combination of generalized Pauli operations. The main result of this section is that even given an arbitrary finite number ($N_E$) of ancilla qudits, a Clifford-restricted Eve (acting with a Clifford gate on the joint system $BE$) cannot implement the optimal attack for all values of $b_{r,s}$, as she cannot simulate the effective channel of Eq.~\eqref{eqn:channelB}.

To see this, consider the Choi dual-state for Eve's Clifford attack $\hat{U} \in \mathcal{C} \ell_d^{\otimes N_E + 1}$
\begin{align}
  \rho_U = \iden_{D} \otimes \hat{U}\ketbra{B^D_{0,0}}\iden_{D} \otimes \hat{U}^\dagger \label{eqn:ChoiU}
\end{align}
where $D = d^{N_E + 1}$. One can easily show that the maximally entangled state for any dimension $D$ can be written as the following sum in the generalized Pauli basis
\begin{align}
  \ketbra{B^D_{0,0}} = \frac{1}{D}\sum_{\substack{r,s,s'\\ s+s'=0~{\rm mod}~D}}^{D-1}\hat{P}^D_{r,s}\otimes\hat{P}^D_{r,s'}.
\end{align}
Putting this into Eq.~\eqref{eqn:ChoiU}, and using the fact the Clifford gates permute the members of the Pauli group, we have that
\begin{align}
  \rho_U = \frac{1}{D}\sum_{\substack{r,s,s'\\ s+s'=0~{\rm mod}~D}}^{D-1}e^{i\phi_{q(r,s')}}\hat{P}^D_{r,s}\otimes\hat{P}^D_{q(r,s')}, \label{eqn:ChoiUfull}
\end{align}
where $q(r,s')$ describes the permutation action of the Clifford gate $\hat{U}$ and $\phi_{q(r,s')}$ describes the associated phase factor.

To determine the Choi dual-state for the effective process on Bob's qudit alone, we take the partial trace with respect to all $N_E$ of Eve's ancilla qubits on both $D$-dimensional subsystems of the Choi dual-state \cite{PAPA}
\begin{align}
  \rho_{U_B} =\frac{1}{d}\sum^{d-1}_{n,m,j,k}\mu(n,m,j,k)\hat{P}^d_{n,m}\otimes\hat{P}^d_{j,k}, \label{eqn:ChoiClifford}
\end{align}
where $\mu(n,m,j,k)$ is either zero or a complex phase (see appendix \ref{app:CG} for details).

The expression for $\rho_{U_B}$ should be compared to the Choi dual-state for the optimal attack channel given by Eq.~\eqref{eqn:local}, which can be rewritten in the Pauli basis as
\begin{align}
  \rho_{AB}= \sum_{r,s}^{d-1}\frac{b_{r,s}}{d}\sum_{\substack{n,k,k'\\ k+k'=0~{\rm mod}~d}}^{d-1}e^{i\phi_{q_{r,s}(n,k')}}\hat{P}^d_{n,k}\otimes\hat{P}^d_{q_{r,s}(n,k')} \label{eqn:ChoiOpt}
\end{align}
where we have used the fact that Pauli operators are themselves Clifford operators and so permute the elements of the Pauli group. Note that each term in the Pauli-random optimal attack channel results in a unique permutation, $q_{r,s}$, of the Pauli basis elements.

Eq.~\eqref{eqn:ChoiClifford} and Eq.~\eqref{eqn:ChoiOpt} are in general inequivalent, as is illustrated by the fact that each of the Pauli basis elements in $\rho_{U_B}$ has a coefficient that has magnitude either $0$ or $1/d$, while in $\rho_{AB}$ the corresponding coefficient can have any complex value with magnitude less than $1/d$ (determined by a sum of elements from the set $\{e^{i\phi_{q_{r,s}(n,k')}}b_{r,s}\}$). Thus, except for specific values of $b_{r,s}$, Eve cannot simulate the optimal attack channel with only a Clifford gate, and so cannot implement the optimal attack.

As a specific example, consider BB84, where the optimal attack channel is given by
\begin{align}
  \nonumber\qp_B(\rho_B) &= (1-Q)^2\iden\rho_B\iden + Q^2\hat{Y}\rho_B\hat{Y} \\ &+ Q(1-Q)\left(\hat{X}\rho_B\hat{X} + \hat{Z}\rho_B\hat{Z}\right), \label{eqn:channelBB84}
\end{align}
with $Q$ the average bit-error rate for all signal-state bases. For $Q = 0,1,0.5$, this effective channel can be implemented using a Clifford gate on $BE$, and we cannot rule out the possibility that restricted Eve can implement the optimal attack. For all other values of $Q$ this is not the case. Note that for $Q = 0,1$ the key rate is not reduced by Eve's interaction.

As Eq.~\eqref{eqn:ChoiClifford} shows, a Clifford attack can be understood as a permutation of the Pauli basis operators of system $B$, along with an acquired phase. For BB84, the two-qubit maximally entangled state can be written in the Pauli basis as
\begin{align}
  \nonumber\ketbra{B^2_{0,0}} = \frac{1}{4}\left(\iden\otimes\iden+\hat{X}\otimes\hat{X}-\hat{Y}\otimes\hat{Y}+\hat{Z}\otimes\hat{Z}\right),
\end{align}
and there are six allowed permutations of the Pauli operators on system $B$, with each element taking one of two phases ($\pm1$). Not all 64 possibilities result in a valid density matrix, but for those that do we calculate the key rate and compare to the worst case of Eq.~\eqref{eqn:channelBB84}, with $Q$ determined from the output state of the Clifford attack channel. As expected, for $Q=0,1,0.5$ the Clifford and worst case key rates are the same. For $Q=0.25,0.75$ (the only other values possible for Clifford attacks), the Clifford attack key rate is finite, while the the worst case key rate is zero.

\subsection{Clifford-Random Channel Restriction}\label{sec:CLifford_restriction}

For the protocols we consider, we have shown that a Pauli-random channel describes the effective channel on Bob's qudit for the optimal eavesdropping attack, and that this cannot be implemented by a Clifford gate on the combined system $BE$. Given these facts, a natural relaxation for the eavesdropper restriction would be to give her the ability to perform Clifford-random channels
\begin{align}
  \mathcal{E}(\rho) = \sum_{g \in C \ell} a_g~\iden^A\otimes\hat{U}_g\rho\iden^A\otimes\hat{U}_g^\dagger,
\end{align}
where $\hat{U}_g$ is an $(N_E+1)$-qudit Clifford gate, and $\sum_g a_g = 1$. Practically, such a channel can be implemented by Eve if she acts with a random Clifford gate on each signal state she intercepts. This requires only the ability to implement Clifford gates, plus a source of classical randomness.

Given such capability, the eavesdropper can directly implement the effective channel of Eq.~\eqref{eqn:channelB}, and \emph{simulate} the statistics of the optimal attack. However, this does not mean that that the eavesdropper can implement the optimal attack that gives her the most information. As such, we wish to quantify how much information a Clifford-random channel restricted Eve can obtain, by determining a lower bound to the secret key rate for this restriction. To do so, we will use the numerical optimization approach developed in Refs.~\cite{Coles:2016aa,Winick:2018aa}, where a computational restriction on the eavesdropper will introduce constraints to the optimization, shrinking the feasible set of states $\rho_{AB}$ describing the optimal attack.

Let us denote the combined state of $ABE$ after the Clifford-random channel as $\rho_{ABE}$, which will in general be a mixed state. We introduce an auxiliary Hilbert space $F$ that is sufficiently large to purify $\rho_{ABE}$ into the pure state $\rho_{ABEF}$. Using the results of Refs.~\cite{Coles:2012aa,Coles:2016aa}, it is straightforward to see that the secret key rate is given by the expression
\begin{align}
  &k(\rho_{ABF}) = r(\rho_{ABF}) - p_{\rm pass}{\rm leak}_{\rm EC}, \label{eqn:keyrate} \\
  \nonumber&r(\rho_{ABF}) = D(\mathcal{G}(\rho_{ABF})||\mathcal{Z}^{R}(\mathcal{G}(\rho_{ABF})))
\end{align}
where $D(\rho||\sigma) = {\rm Tr}[\rho(\log\rho-\log\sigma)]$ is the quantum relative entropy.

As described in detail in Ref.~\cite{Winick:2018aa}, $\mathcal{G}(.)$ is a completely positive trace-nonincreasing quantum channel that describes the measurements, public announcements, and post selection performed by Alice and Bob, with $p_{\rm pass}$ the probability that the post selection is successful. The term ${\rm leak}_{\rm EC}$ describes the information leaked (and available to the eavesdropper) during the error correction used to generate a secure key, and $\mathcal{Z}^{R}(.)$ is a pinching channel over the Hilbert space $R$ that records the results of Alice's key-map
\begin{align}
  \mathcal{Z}^{R}(\rho) = \sum_i \ketbra{i}^{R}\rho\ketbra{i}^{R},
\end{align}
where $\{\ket{i}_{R}\}$ is an orthonormal basis for $R$.

To find a lower bound for the key rate, we would solve the nonlinear semi-definite program (SDP), defined in Ref.~\cite{Winick:2018aa}, that finds the state $\rho^*_{ABF}$ that minimizes $r(\rho_{ABF})$ from Eq.~\eqref{eqn:keyrate}, taking into account constraints on the final state $\rho_{ABF}$ that come from channel reconciliation, such as the quantum bit error rate. Unfortunately, without knowledge of the nature of $F$, we cannot calculate Eq.~\eqref{eqn:keyrate}, and thus cannot evaluate the cost function of our SDP.

However, using the monotonicity of the quantum relative entropy under completely positive and trace preserving maps, we have that
\begin{align}
  \nonumber D(\mathcal{G}(\rho^*_{ABF})||\mathcal{Z}^{R}(\mathcal{G}(\rho^*_{ABF}))) \geq D(\mathcal{G}(\rho^*_{AB})||\mathcal{Z}^{R}(\mathcal{G}(\rho^*_{AB})))
\end{align}
where $\rho^*_{AB} = {\rm Tr}_F\left[\rho^*_{ABF}\right]$. Using this, clearly
\begin{align}
  r(\rho^*_{ABF}) \geq r(\rho^*_{AB}) \geq r(\rho'_{AB}),
\end{align}
with $\rho'_{AB}$ the optimal solution for a key rate SDP over states in $AB$. It is straightforward to modify the SDP of Ref.~\cite{Winick:2018aa} to include constraints from restricted Eve (see appendices \ref{app:SDP} and \ref{sec:REsetup} for details), and thus we can numerically find a \emph{lower bound} to the key rate for the optimal Clifford-random channel attack.

In practice, numerical SDP solvers are not able to return the exact optimal solution $\rho'_{AB}$. Solving the primal problem of the SDP will in general give an upper bound to the optimum, so instead we solve the dual problem, which gives a lower bound. This guarantees that $r(\rho'_{AB})$ is a lower bound to the secret key rate. To solve the dual problem we use the convex approximation to the quantum relative entropy developed in \cite{Fawzi:2018aa}, which integrates with CVX, a Matlab package for specifying and solving convex problems \cite{cvx,gb08}. In appendix \ref{app:sec} we discuss how to obtain a \emph{secure} lower bound using either the techniques from Ref.~\cite{Fawzi:2018aa} or from Ref.~\cite{Winick:2018aa}, and present numerical evidence that our approximate lower bounds are accurate to high precision.

\subsection{Clifford-Random Channel Restricted BB84}

As an example of our numerical approach, we consider prepare-and-measure BB84, with the modifcation that we allow the initial state to be any state of the form
\begin{align}
  \ket{\Psi_{I}}_{AA'} = \sqrt{b}\ket{00} + \sqrt{1-b}\ket{11}, \label{eqn:psi_in}
\end{align}
where $0\leq b \leq 1$. After Alice prepares this initial state, she sends the second register $A'$ to Bob via the insecure quantum channel that Eve has access to, and we call the register after the transmission $B$. For $b\neq1/2$ these initial states lead to BB84 protocols that we refer to as asymmetric.

The protocol we consider is outlined as follows (details can be found in appendix \ref{app:numcalc}):
\begin{itemize}
  \item Alice and Bob each independently measure in the Z-basis or X-basis, with probability $p$ and $1-p$ respectively.
  \item Alice's key-map maps the $+1$ measurement outcome to the character 0, and a $-1$ outcome to the character 1, i.e.
  \begin{align}
    \{\ket{0},\ket{+}\}\rightarrow 0~~~~\{\ket{1},\ket{-}\}\rightarrow 1
  \end{align}
  where $\ket{\pm} \propto \ket{0} \pm \ket{1}$ are the eigenstates of $\hat{X}$.
  \item Alice and Bob perform channel reconciliation, which adds one of two sets of constraints to the SDP, both of which can be calculated from asymptotic measurement statistics.
  \begin{enumerate}
    \item Coarse-grained constraints:
    \begin{align}
      {\rm Tr}\left[\rho_{AB}\hat{E}_{Z/X}\right] = \gamma_{z/x}, \label{eqn:gamma_zx}
    \end{align}
    where
    \begin{align}
      &\hat{E}_{Z} = \ketbra{01} + \ketbra{10}, \label{eqn:Ez}\\
      &\hat{E}_{X} = \ketbra{+-} + \ketbra{-+}. \label{eqn:Ex}
    \end{align}

    \item Fine-grained constraints:
    \begin{align}
      {\rm Tr}\left[\rho_{AB}\hat{E}_{ij}\right] = \gamma_{ij}, \label{eqn:gamma_fi}
    \end{align}
    where
    \begin{align}
      \hat{E}_{ij} = \hat{M}_i\otimes\hat{M}_j
    \end{align}
    with $\hat{M}_i\in\{\ketbra{0},\ketbra{1},\ketbra{+},\ketbra{-}\}$
  \end{enumerate}
  \item Further constraints to the SDP come from the complete knowledge of Alice's reduced state, to which Eve does not have access, and from our knowledge of Eve's restriction.
\end{itemize}
We simulate data to calculate the constraints and the error correction cost ${\rm leak}_{\rm EC}$ using a depolarizing channel with depolarizing probability $\epsilon$ (see appendix \ref{app:EC} for further details). For our input state on $AA'$, this produces an output state on $AB$ given by
\begin{align}
  \qp_{\rm dep}(\rho_{AA'}) = \left(1-\epsilon\right)\rho_{AA'} + \epsilon{\rm Tr}_B\left[\rho_{AA'}\right]\otimes \frac{1}{2}\iden^B,
\end{align}
where $\rho_{AA'}$ is the pure state of Eq.~\eqref{eqn:psi_in}. We will alternatively refer to $\epsilon$ as the channel error, as it is the error added to the state by the insecure channel between Alice and Bob.

\begin{figure}[ht]
  \includegraphics[width=0.75\columnwidth]{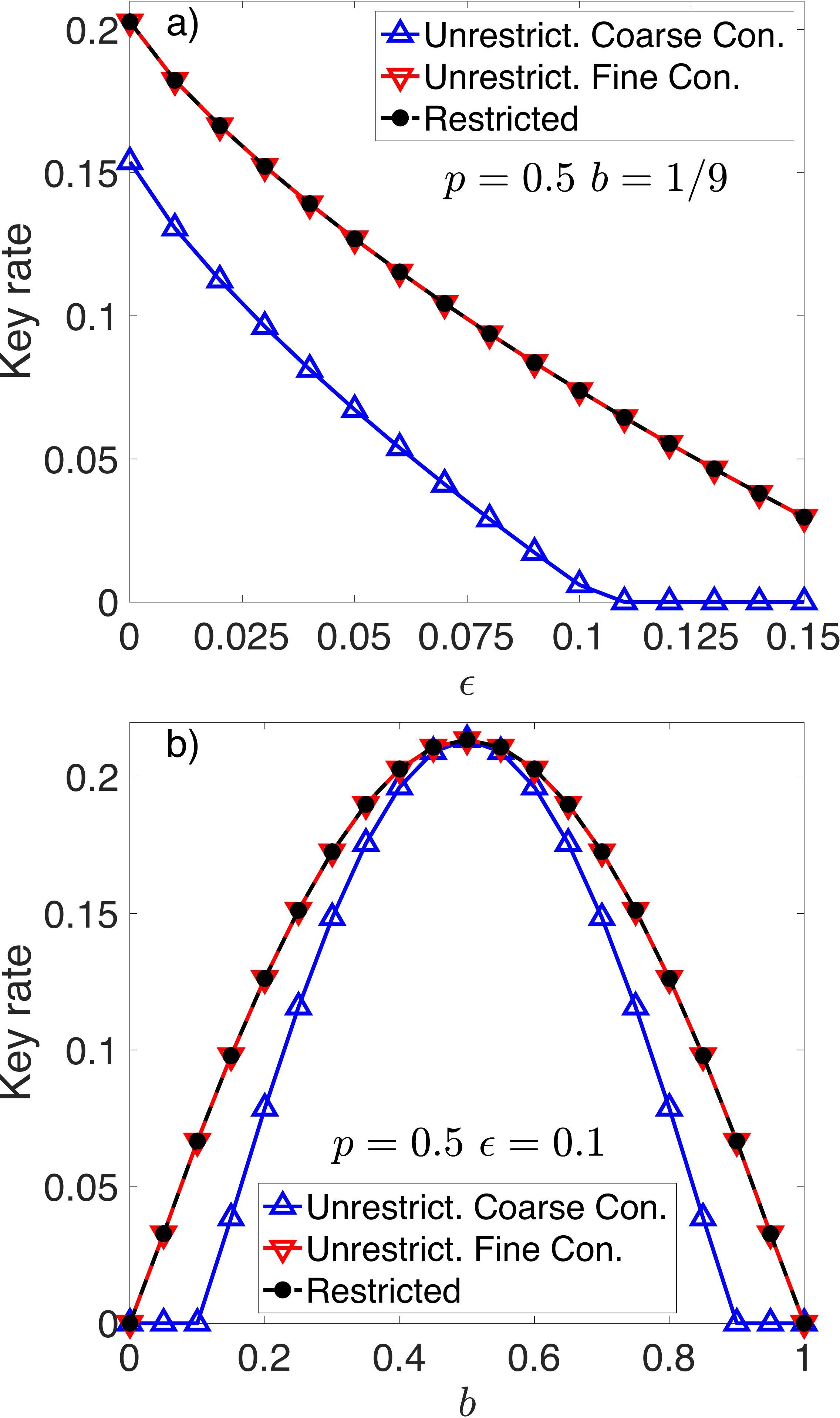}
  \caption{Secret key rate lower bound function as a function of {\bf a)} channel error $\epsilon$ and {\bf b)} asymmetry parameter $b$,  with or without a restricted eavesdropper, and with either coarse or fine grained constraints are considered. For the restricted eavesdropper the key rate is the same for both sets of constraints. Fixed parameters indicated on each subplot.}
  \label{fig:RE}
\end{figure}

The results of our numerical calculations of key rate lower bounds for a Clifford-channel restricted Eve are shown in Fig.~\ref{fig:RE}. These are compared with the \emph{key rates} for an unrestricted Eve (we solve both the primal and dual problem, and find a duality gap much smaller than our reported key rate). Fig.~\ref{fig:RE}a) shows the key rate lower bound for an asymmetric initial state ($b=1/9$) as a function of the channel error $\epsilon$, while Fig.~\ref{fig:RE}b) varies the asymmetry parameter $b$, and keeps the channel error fixed to $\epsilon = 0.1$. In both cases the probability that either Alice or Bob measure in the Z-basis is $p=0.5$.

As both plots show, the comparison between the restricted and unrestricted key rates depends strongly on the constraints used in the SDP, i.e.~in how much characterization Alice and Bob do of the insecure channel connecting them. For the coarse-grained constraints the restricted key rate lower bound can be significantly larger than the unrestricted key rate (depending on the asymmetry), while for the fine-grained constraints they are identical (note that the restricted key rate lower bound is the same regardless of the constraints). This result is unsurprising, as from the perspective of the SDP optimization, restricting Eve is equivalent to adding additional constraints. It turns out that for BB84 the additional constraints coming from a restricted Eve can also be determined by fine-grained channel reconciliation. In fact, we can determine that it is the constraint
\begin{align}
  &\nonumber\gamma_{IZ} = \left<\hat{I}^A\otimes\hat{Z}^B\right>_{\rho_{AB}} =\sum_{n,k=0}^1(-1)^k\bra{n,k}\rho_{AB}\ket{n,k},\\
  &= \gamma_{00} - \gamma_{01} + \gamma_{10} - \gamma_{11}
\end{align}
provided by both restricted Eve and the fine-grained constraints that is responsible.

\section{Discussion}
\label{sec:Conclusions}

In this paper, we have studied the role of computational capabilities of the eavesdropper on secret key rates achievable for a QKD protocol. Specifically, for discrete variable protocols with $d$-MUBs and Weyl-Heisenberg symmetry, we found that the optimal attack for the eavesdropper cannot be implemented by a single Clifford gate with even an arbitrary number of ancilla qudits given to the eavesdropper. In the specific case of BB84, finite key rates are possible for a Clifford-gate restricted eavesdropper when the worst case attack has zero key rate.

We then considered Clifford-random channels, and derived a way to use the numerical approach of Ref.~\cite{Winick:2018aa} to find a lower bound for eavesdropper attacks with a total final state that is mixed. Again for BB84, we calculated the secret key rate with a Clifford-random channel restricted eavesdropper by modifying the SDP to incorporate the restriction. To do so, we have developed several reductions to the QKD numerical SDP problem to make it computationally tractable, generalizations of which may find use in other numerical studies of QKD protocols (see Ref.~\cite{Lin:2019aa} and appendix \ref{app:numcalc} for further details). We found that the difference between the worst case Clifford-random channel restricted key rate and the optimal attack key rate depends on the information obtained during channel reconciliation.

For BB84 we found that a Clifford-random channel restriction does not reveal any information about the channel not available from full channel reconciliation, and there is no key rate improvement for a restricted eavesdropper. However, we have only calculated a lower bound to the key rate, due to our inability to describe the purifying system $F$. To calculate the key rate, or bound it more tightly, future research is needed to develop methods for calculating secret key rates with mixed final states (where the problem cannot be cast in terms of the quantum relative entropy), or that incorporate the purifying auxiliary system \cite{Haseler09,Killoran12}.

Further, while we have not shown an improved key rate lower bound for Clifford-random channel restricted BB84, we have analytically shown an improved key rate for Clifford-gate restricted BB84. Thus, we anticipate that numerical approaches similar to ours will find improved key rates for other restrictions, and other protocols, especially those with systems in larger Hilbert spaces. We intend our work to serve as the initiation point for the general study of QKD with computational restrictions. This will be especially relevant in the near term, where QKD technological development is likely to outpace the development of universal quantum computers.

We have studied restrictions involving Clifford gates, as these are most likely to be fault-tolerantly implementable in the near future (e.g.~on the surface code \cite{Fowler:2012aa}), and can be implemented transversally i.e., in a parallel fashion. Further, they are analogous to Gaussian operations in the continuous variable (CV) case, which can be implemented with only linear optics, and hence provide a reasonable model for restrictions on the eavesdropper for CV systems. Future work will seek to generalize the study of computational restrictions to such CV-QKD protocols, and to implementations of DV-QKD using weak coherent states.

\acknowledgements
L.G., D.B.~and H.K.~were supported by the Office of Naval Research program Communications and Networking with Quantum Operationally-Secure Technology for Maritime Deployment (CONQUEST), awarded under contract number N00014-16-C-2069.

\appendix

\section{Derivation of the Clifford Gate Effective Channel}
\label{app:CG}

The generalized Bell states for two qudits are
\begin{align}
  \ket{B^d_{r,s}} = \frac{1}{\sqrt{d}}\sum_{k=0}^{d-1}e^{\frac{2i\pi}{d}ks}\ket{k}\ket{k+r},
\end{align}
and the generalized Pauli matrices of dimension $d$ are
\begin{align}
  \hat{P}^d_{r,s} = \sum_{k=0}^{d-1}e^{\frac{2i\pi}{d}ks}\ketbra{k+r}{k}.
\end{align}
From Eq.~\eqref{eqn:ChoiUfull}, deriving the Choi state for the effective process on Bob's qudit alone is done by taking the partial trace of Eve's $N_E$ ancilla qubits as follows
\begin{align}
  &\nonumber\rho_{U_B} \\&= \frac{1}{D}\sum_{\substack{r,s,s'\\ s+s'=0}}^{D-1}e^{i\phi_{q(r,s')}}{\rm Tr}_{1...N_E}\left[\hat{P}^D_{r,s}\right]\otimes{\rm Tr}_{1...N_E}\left[\hat{P}^D_{q(r,s')}\right] \nonumber \\
  &=\frac{1}{d}\sum^{d-1}_{n,m,j,k}\mu(n,m,j,k)\hat{P}^d_{n,m}\otimes\hat{P}^d_{j,k},
\end{align}
where for brevity, when we write $s+s'=0$ it should be understood that this is addition modulo $D$. To arrive at the last line we have used the fact that
\begin{align}
  {\rm Tr}\left[\hat{P}^d_{r,s}\right] = d\delta_{r,0}\delta_{s,0},
\end{align}
and that a $d^{N_E}$-dimensional Pauli operator is a tensor product of $N_E$ $d$-dimensional Pauli operators. The complex weight function $\mu(n,m,j,k)$ is zero unless
\begin{align}
  \hat{P}^D_{r,s} =\iden_d^{\otimes N_E} \otimes \hat{P}^d_{n,m}~{\rm and}~\hat{P}^D_{q(r,s')} = \iden_d^{\otimes N_E} \otimes \hat{P}^d_{j,k}
\end{align}
can be satisfied simultaneously for some valid $r$, $s$, and $s'$, in which case it takes the value of $e^{i\phi_{q(r,s')}}$.

\section{Semi-definite Program Setup for BB84}
\label{app:SDP}

With no restriction on the eavesdropper, there is no need to introduce the auxilliary system $F$, and we can directly solve an SDP for $\rho_{AB}$ \cite{Winick:2018aa}. We solve the SDP
\begin{gather*}
  \min_{\rho_{AB}}\Big(p^2D(\rho_{AB}||\mathcal{Z}_Z^A(\rho_{AB})) + (1-p)^2D(\rho_{AB}||\mathcal{Z}_X^A(\rho_{AB}))\Big)\\
  \rho_{AB} \geq 0,~{\rm Tr}\left(\rho_{AB}\right) = 1\\
  {\rm Tr}\left(\rho_{AB}\hat{E}_{Z/X}\right) = \gamma_{z/x}~{\rm OR}~{\rm Tr}\left(\rho_{AB}\hat{E}_{ij}\right) = \gamma_{ij}\\
  \left<\hat{X}^A\right>_{\rho_{AB}} = \left<\hat{Y}^A\right>_{\rho_{AB}} =0 ,~\left<\hat{Z}^A\right>_{\rho_{AB}} = 2b^2-1
\end{gather*}
where the minimization is over $\rho_{AB} \in B(\mathcal{H}_{AB})$. The second line of constraints implements either the coarse- or fine-grained constraints from channel reconciliation. Here $\mathcal{Z}_\mu^{K}(.)$ is the pinching channel on Hilbert space $K$ in the basis of the operator $\hat{\mu}$. The derivation of this SDP can be found in Ref.~\cite{Winick:2018aa}, except for the separation of the objective function into two parts, which can be found in appendix \ref{app:numcalc}. This separation enables all operations to be on the $AB$ Hilbert space alone. The last line of constraints comes from the complete knowledge of the reduced state of Alice, where we have used the shorthand
\begin{align}
  \left<\hat{\mu}^A\right>_{\rho_{AB}} = {\rm Tr}\left(\rho_{AB}\hat{\mu}\otimes\iden\right),
\end{align}
for compactness.

A restriction on Eve's computational power places additional constraints in this optimization problem, which reduces the set of feasible $\rho_{AB}$. As such we can augment the SDP written above with additional constraints on $\rho_{AB}$ that encode how the set of operations the eavesdropper can perform is restricted. However, for our Clifford-random channel restriction we find it simpler to introduce a parameterization of $\rho_{AB}$ in the feasible set in a smart choice of matrix basis, as it is not straightforward to construct additional constraints of the standard form used in \cite{Winick:2018aa}. Some of the coefficients in this parameterization are constrained optimization parameters, which we label $C_{\mu}^j$.

For a Clifford-random channel restricted eavesdropper, we solve the SDP
\begin{gather*}
  \min_{C_{\mu}^j\in\mathbb{R}} r\left(\rho_{AB}\left[C_{\mu}^j\right]\right)  \\
  \rho_{AB} \geq 0\\
  {\rm Tr}\left(\rho_{AB}\hat{E}_{Z/X}\right) = \gamma_{z/x},~{\rm OR}~{\rm Tr}\left(\rho_{AB}\hat{E}_{ij}\right) = \gamma_{ij}\\
  \sum_{j,\mu}C_{\mu}^j = 3,~\sum_{\mu}C_{\mu}^j \leq 1~\forall j,~0\leq C_{\mu}^j \leq 1
\end{gather*}
where $\rho_{AB}\left[C_{\mu}^j\right]$ is the state written in our chosen matrix basis, with free parameters $C_{\mu}^j$. For compactness of notation we have suppressed the functional dependence of $\rho_{AB}$ on the free parameters $C_{\mu}^j$ below the first line. The function $r\left(\rho_{AB}\left[C_{\mu}^j\right]\right)$ is the same as in the unrestricted SDP, but with $\rho_{AB}$ replaced by $\rho_{AB}\left[C_{\mu}^j\right]$. The optimization domain becomes the real numbers for the free parameters $C_{\mu}^j$.

The parameterized $\rho_{AB}\left[C_{\mu}^j\right]$ lies within a subset of the Hermitian, trace one matrices, defined by additional constraints coming from the eavesdropper's restriction to Clifford-random channels. The complete knowledge of Alice's reduced state is also encoded in this parameterization, so those constraints are not added to the SDP as they are for the unrestricted case. A complete description of the parameterization of $\rho_{AB}\left[C_{\mu}^j\right]$ and its derivation can be found in appendix \ref{sec:REsetup}.

Both SDPs require optimizations over density matrices within the $AB$ Hilbert space alone. However, unlike in the standard problem setup \cite{Winick:2018aa}, we have derived an SDP where the operations performed on $\rho_{AB}$ do not require dilation to a larger Hilbert space that includes descriptions of measurement outcomes, public announcements, and post selection. For instance, the pinching channels are on the Hilbert space $A$ instead of $R$. This greatly improves the numerical performance of the optimization. A detailed description of how we have reduced the operational Hilbert space dimension of the secret key rate function can be found in appendix \ref{app:numcalc}, and is related to techniques used in \cite{Lin:2019aa}.

\section{Restricted Eavesdropper for Qubit Systems}
\label{sec:REsetup}

\subsection{General Setup}

In a prepare-and-measure setup, via a source replacement scheme we can describe the initial state of Alice and Bob as
\begin{align}
  \ket{\Psi_{I}}_{AA'} = \sum_n\sqrt{p_n}\ket{n}\ket{\phi_n},
\end{align}
where Alice records the choice of the state she prepares in the register $A$ and sends the state in register $A'$ to Bob which becomes the system $B$ after the channel controlled by Eve. The full initial state of Alice, Bob and Eve is
\begin{align}
  \ket{\Psi_{I}}_{AA'E} = \ket{\Psi_{I}}_{AA'}\ket{0}^{\otimes N_E}.
\end{align}
where assuming Eve's initial state is pure, we can with complete generality define it to be the product state of all Eve's ancilla qubits in the state $\ket{0}$. Eve can act only on the $A'$ half of the initial $AA'$ state, as well as her own state. Assuming Eve implements the unitary $\hat{U} \in B\left(\mathcal{H}_{A'}\otimes\mathcal{H}_E\right)$, which is a general description for Eve's attack under the i.i.d.~assumption we have made, then the final state of $ABE$ is given by
\begin{align}
  \rho_{ABE} = \hat{U}\ketbra{\Psi_{I}}_{AA'E}\hat{U}^\dagger.
\end{align}
The state of $AB$ alone is $\rho_{AB} = {\rm Tr}_E\left(\rho_{ABE}\right)$, which we parameterize as
\begin{align}
  \rho_{AB} = \sum_i \gamma_i \hat{\Gamma}_i + \sum_j \omega_j \hat{\Omega}_j.
\end{align}
Here $\hat{\Gamma}_i$ are the operators whose expectation value with the state $\rho_{AB}$ is known, either through the complete knowledge of the state of $A$, $\rho_A = {\rm Tr}_{AE}\left(\rho_{ABE}\right)$, or from parameter estimation. The set $\hat{\Omega}_j$ complete an orthonormal basis for operator space, and have unknown expectation values given by
\begin{align}
  \nonumber\omega_j &= {\rm Tr}\left(\rho_{AB}\hat{\Omega}_j\right) = {\rm Tr}\left(\hat{U}\ketbra{\Psi_{I}}_{ABE}\hat{U}^\dagger\hat{\Omega}_j\right) \\
  &= \bra{\Psi_{I}}_{ABE}\hat{U}^\dagger\hat{\Omega}_j \hat{U}\ket{\Psi_{I}}_{ABE} \equiv \left<\hat{U}^\dagger\hat{\Omega}_j \hat{U}\right>, \label{eqn:omj}
\end{align}
where the last expression is a notational definition we use for convenience.

\subsection{Restricting to Clifford Operations}

Let us assume that Eve can only implement Clifford gates on the combined system $BE$. Then it is smart to choose $\hat{\Gamma}_i$ and $\hat{\Omega}_j$ to be Pauli operators. In Eq.~\eqref{eqn:omj} we consider the full system $ABE$, and so append the identity operator acting on Eve's system to $\hat{\Gamma}_i$ and $\hat{\Omega}_j$, such that we set
\begin{align}
  &\hat{\Gamma}_i = \hat{P}_i^{A}\otimes\hat{P}_i^{B}\otimes\hat{\mathbb{I}}^{\otimes N_E}, \\
  &\hat{\Omega}_j = \hat{P}_j^{A}\otimes\hat{P}_j^{B}\otimes\hat{\mathbb{I}}^{\otimes N_E},
\end{align}
with $\hat{P}_j^{A}, \hat{P}_j^{B} \in \frac{1}{\sqrt{2}}\{\hat{\mathbb{I}},\hat{X},\hat{Y},\hat{Z}\}$ normalized single-qubit Pauli operators. The restriction to Clifford operations is such that Eve can only implement a unitary $\hat{U}_g\in C \ell \left(\mathcal{H}_{A'}\otimes\mathcal{H}_E\right)$, and the operation on the full Hilbert space $ABE$ is given by
\begin{align}
  \hat{U} = \hat{\mathbb{I}}^{A}\otimes\hat{U}_g.
\end{align}

As Clifford operations map Pauli matrices to other Pauli matrices, we see that
\begin{align}
  \omega_j = \left<\left(\hat{\mathbb{I}}^{A}\otimes\hat{U}^\dagger_g\right)\hat{\Omega}_j\left( \hat{\mathbb{I}}^{A}\otimes\hat{U}_g\right)\right> = \left<\pm\hat{\Omega}^{(g)}_j\right>,
\end{align}
where $\hat{\Omega}^{(g)}_j = \hat{P}_j^{A}\otimes\hat{P}^{(g)}_j$ with $\hat{P}^{(g)}_j = \pm\hat{U}^\dagger_g\left(\hat{P}_j^{B}\otimes\hat{\mathbb{I}}^{\otimes N_E}\right)\hat{U}_g$ an $N_E+1$ qubit Pauli operator. Note that the $\pm$ sign in the expecation value is meant to imply that depending on the Clifford gate $\hat{U}_g$, we may have either $+\hat{\Omega}^{(g)}_j$ or $-\hat{\Omega}^{(g)}_j$.

Extending to Clifford-random channels, Eve's action is generically described by
\begin{align}
  \rho_{ABE} = \sum_{g \in C \ell} a_g~\iden^A\otimes\hat{U}_g\ketbra{\Psi_{I}}_{AA'E}\iden^A\otimes\hat{U}_g^\dagger,
\end{align}
with $\sum_g a_g = 1$. Then for the unknown parameters $\omega_j$ we have
\begin{align}
  \omega_j = \sum_g a_g \left<\left(\hat{\mathbb{I}}^{A}\otimes\hat{U}^\dagger_g\right)\hat{\Omega}_j\left( \hat{\mathbb{I}}^{A}\otimes\hat{U}_g\right)\right>,
\end{align}
where, as a reminder $\hat{\Omega}_j$ is a two-qubit Pauli operator on $AB$ (dilated to account for $E$). To make this more explicit, we define
\begin{align}
    \nonumber\omega_j^k &= \sum_g a_g \left<\left(\hat{\mathbb{I}}^{A}\otimes\hat{U}^\dagger_g\right)\hat{P}^A_k\otimes\hat{P}^B_j\otimes\hat{\mathbb{I}}^{\otimes N_E}\left( \hat{\mathbb{I}}^{A}\otimes\hat{U}_g\right)\right> \\ &= \sum_g a_g \left<\hat{P}^A_k\otimes\hat{P}^B_{j,g}\otimes\hat{P}^{E}_{j,g}\right>, \label{eqn:ojk}
\end{align}
where we have defined
\begin{align}
  \hat{P}^B_{j,g}\otimes\hat{P}^{E}_{j,g} = \hat{U}^\dagger_g\left(\hat{P}^B_j\otimes\hat{\mathbb{I}}^{\otimes N_E}\right)\hat{U}_g,
\end{align}
with $\hat{P}_{j,g}^E$ a $N_E$-qubit Pauli matrix that is \emph{not} normalized. From here, the first key observation is that since Eve's initial state is $\ket{0}^{\otimes N_E}$, then only those $N_E$-qubit Pauli operators contained within the set $\hat{P}^{E}_{j,g}\in\{\iden,\hat{Z}\}^{\otimes N_E}$ have a nonzero expectation value with Eve's initial state. Thus we have that
\begin{align}
  \left<\hat{P}^{E}_{j,g}\right> = 0~\text{or}~\pm 1 \label{eqn:Evals}
\end{align}
and we can rewrite Eq.~\eqref{eqn:ojk} as
\begin{align}
  \nonumber&\sum_g a_g \left<\hat{P}^A_k\otimes\hat{P}^B_{j,g}\otimes\hat{P}^{E}_{j,g}\right> = \sum_g a_g \left<\hat{P}^A_k\otimes\hat{P}^B_{j,g}\right>\left<\hat{P}^{E}_{j,g}\right> \\
  &=  \sum_{g^j_+} a_{g^j_+} \left<\hat{P}^A_k\otimes\hat{P}^B_{j,g}\right> - \sum_{g^j_-} a_{g^j_-} \left<\hat{P}^A_k\otimes\hat{P}^B_{j,g}\right>. \label{eqn:Eex}
\end{align}
The last expression not only splits the summation into two parts, but drops all $a_g$ with $\left<\hat{P}^{E}_{j,g}\right>=0$, and so the sum of the remaining $a_g$ may be less than one.

Further, we know that for $\hat{P}_j^B \neq \iden/\sqrt{2}$, the action of $\hat{U}_g$ can only transform $\hat{P}_j^B$ such that $\hat{P}^B_{j,g} \in \frac{1}{\sqrt{2}}\{\hat{X},\hat{Y},\hat{Z}\}$. This allows us to split Eq.~\eqref{eqn:ojk} into 6 separate sums, distinguished by $\hat{P}^B_{j,g}$ and Eq.~\eqref{eqn:Eex}
\begin{align}
  \nonumber\sqrt{2}\omega_j^k &= \sum_{g^j_X} a_{g^j_X} \left<\hat{P}^A_k\otimes\hat{X}^B\right> + \sum_{g^j_Y} a_{g^j_Y} \left<\hat{P}^A_k\otimes\hat{Y}^B\right> \\
  \nonumber&+ \sum_{g^j_Z} a_{g^j_Z} \left<\hat{P}^A_k\otimes\hat{Z}^B\right> - \sum_{g^j_{-X}} a_{g^j_{-X}} \left<\hat{P}^A_k\otimes\hat{X}^B\right> \\
  &- \sum_{g^j_{-Y}} a_{g^j_{-Y}} \left<\hat{P}^A_k\otimes\hat{Y}^B\right> - \sum_{g^j_{-Z}} a_{g^j_{-Z}} \left<\hat{P}^A_k\otimes\hat{Z}^B\right>, \label{eqn:omkj2}
\end{align}
where now the summation index $g^j_\mu$, for $\hat{\mu} \in \{\hat{X},\hat{Y},\hat{Z}\}$, runs over all $(N_E+1)$-qubit Clifford gates that transform $\hat{P}_{j}^B$ of Eq.~\eqref{eqn:ojk} such that $\sqrt{2}\hat{P}_{j}^B \otimes \iden^{\otimes N_E} \longrightarrow \hat{\mu}\otimes\hat{P}^{E}_g$, with
\begin{align}
  \nonumber& \hat{\mu} \in \{\hat{X},\hat{Y},\hat{Z}\}~{\rm and}~\hat{P}^{E}_g \in\{\iden,\hat{Z}\}^{\otimes N_E}~s.t.~\left<\hat{P}^{E}_g\right> = 1 \\
  &\nonumber {\rm or} \\
  \nonumber&\hat{\mu} \in -\{\hat{X},\hat{Y},\hat{Z}\}~{\rm and}~\hat{P}^{E}_g\in\{\iden,\hat{Z}\}^{\otimes N_E}~s.t.~\left<\hat{P}^{E}_g\right> = -1
\end{align}
which gives a $+$ sign to the coefficient $a_{g^j_\mu}$, and the summation index $g^j_{-\mu}$ runs over all $(N_E+1)$-qubit Clifford gates that transform $\hat{P}_{j}^B$ as before but with
\begin{align}
  \nonumber&\hat{\mu}\in -\{\hat{X},\hat{Y},\hat{Z}\}~{\rm and}~\hat{P}^{E}_g\in\{\iden,\hat{Z}\}^{\otimes N_E}~s.t.~\left<\hat{P}^{E}_g\right> = 1 \\
  &\nonumber {\rm or} \\
  \nonumber&\hat{\mu}\in \{\hat{X},\hat{Y},\hat{Z}\}~{\rm and}~\hat{P}^{E}_g\in\{\iden,\hat{Z}\}^{\otimes N_E}~s.t.~\left<\hat{P}^{E}_g\right> = -1
\end{align}
which gives a $-$ sign to the coefficient $a_{g^j_{-\mu}}$.

We have only considered $j = 1,2,3$ which corresponds to $\hat{P}^B_{j} \in \frac{1}{\sqrt{2}}\{\hat{X},\hat{Y},\hat{Z}\}$, as for $\hat{P}_0^B = \iden/\sqrt{2}$ the expressions for $\omega_0^k$ are trivially given by
\begin{align}
    \nonumber\omega_0^k &= \frac{1}{\sqrt{2}}\sum_g a_g \left<\left(\hat{\mathbb{I}}^{A}\otimes\hat{U}_g\right)\hat{P}^A_k\otimes\iden^B\otimes\iden^{\otimes N_E}\left( \hat{\mathbb{I}}^{A}\otimes\hat{U}_g^\dagger\right)\right> \\
     &= \frac{1}{\sqrt{2}}\bra{\Psi_I}_{AA'}\hat{P}^A_k\otimes\iden^B\ket{\Psi_I}_{AA'},
\end{align}
and describe the initial state of $A$.

The summations in Eq.~\eqref{eqn:omkj2} separate the Clifford group into disjoint subsets defined by the action of its elements on a Pauli operator $\hat{P}_{j}^B$, and therefore the sets of free parameters $\{a_{g_\nu^j}\}$ are disjoint for a given $j$. For each summation, we can therefore define
\begin{align}
    C^j_\nu = \sum_{g^j_\nu} a_{g^j_\nu}, \label{eqn:Ccoeff}
\end{align}
and rewrite Eq.~\eqref{eqn:omkj2} as
\begin{align}
  \nonumber\sqrt{2}\omega_j^k &= C^j_X \left<\hat{P}^A_k\otimes\hat{X}^B\right> + C^j_Y\left<\hat{P}^A_k\otimes\hat{Y}^B\right> \\
  \nonumber&+ C^j_Z \left<\hat{P}^A_k\otimes\hat{Z}^B\right> - C^j_{-X} \left<\hat{P}^A_k\otimes\hat{X}^B\right> \\
  &- C^j_{-Y} \left<\hat{P}^A_k\otimes\hat{Y}^B\right> - C^j_{-Z} \left<\hat{P}^A_k\otimes\hat{Z}^B\right>. \label{eqn:form2}
\end{align}
By properties of the $a_g$, the parameters $C^j_\nu$ satisfy $0\leq C^j_\nu\leq 1$ and $\sum_{\nu}C^j_\nu \leq 1~\forall j$. The normalization is an inequality as not all $a_g$ will survive Eq.~\eqref{eqn:Eex}.

For each $j$, the parameters $C^j_\nu$ are independent of one another as they are sums over disjoint sets of the Clifford group. However, for different $j$ the $C^j_\nu$ are not guaranteed to be independent of one another, as the separation of the Clifford group into sets defined by the action of its elements on a Pauli operator $\hat{P}_{j}^B\otimes\iden^{\otimes N_E}$ depends strongly on the specific $\hat{P}_{j}^B$. Considering all $j$ gives a total of 18 $C^j_\nu$, and we have that $\sum_{j,\nu}C^j_\nu \leq 3$. This normalization inequality is saturated when the Clifford-random channel consists of Clifford gates that act nontrivially for all $\hat{P}_{j}^B\otimes\iden^{\otimes N_E}$, i.e.~never take the value of zero in Eq.~\eqref{eqn:Evals}.

The second key observation of our derivation is that for the purposes of searching for an optimal attack, we can treat all $C^j_\nu$ as independent, up to their normalization, as such a search space will necessarily contain all valid $C^j_\nu$ for a Clifford-random channel attack. Thus, the optimal attack found in this enlarged search space will produce a key rate that is guaranteed to be a lower bound for the worst case key rate due to a Clifford-random channel attack.

We further assert that \emph{all} valid density matrices in the search space with $C^j_\nu$ independent correspond to a valid Clifford-random channel attack. We cannot prove that this is the case, but we give a strong argument in what follows. To see that this is likely the case, we contend that there are enough ($N_E+1$)-Clifford gates that it is possible to select 18 of them such that $C^j_\nu = a_{g_\nu^j}$ are all independent. Other Clifford gates added to the channel can increase the value of one or more $C^j_\nu$. In the case where more than one $C^j_\nu$ have their value increased, then the sum of all $C^j_\nu$ can exceed unity. This effect, plus the 18 gates giving independent control is sufficient to generate any value of the free parameters $C^j_\nu$ such that $0\leq C^j_\nu\leq 1$ , $\sum_{\nu}C^j_\nu \leq 1~\forall j$, $\sum_{j,\nu}C^j_\nu \leq 3$, and $\rho_{AB} \geq 0$. We note that it is possible to select values of $C^j_\nu$ that do not correspond to a valid Clifford-random attack, but only if one of the constraints listed previously is violated.

The selection of 18 Clifford gates for independent $C^j_\nu$ implies that each of these gates acts nontrivially (nonzero expectation value in Eq.~\eqref{eqn:Evals}) on only one of the three possible initial operators $\hat{P}_j^B\otimes\iden^{\otimes N_E}$. There are three sets of six that each act nontrivially on the same initial operator, and within these sets, each Clifford gate maps the initial Pauli operator to one of the six distinct final Pauli operators (including the $\pm1$ phase). The selection of 18 such Clifford gates is possible because the elements of the Clifford group realize all permutations of the Pauli group (with a $\pm1$ phase) that preserve its commutation structure, and we are considering the ($N_E+1$)-Clifford group, with $N_E$ an arbitrarily large, but finite, number.

\subsection{Application to BB84}
\label{subsec:BB84}

Consider asymmetric BB84, with the initial state of Alice and Bob given by
\begin{align}
  \ket{\Psi_{I}}_{AA'} = \sqrt{b}\ket{00} + \sqrt{1-b}\ket{11},
\end{align}
where $0 \leq b\leq1$, as used in the main text. From this we can work out the expressions for $\omega_j^k$ from Eq.~\eqref{eqn:form2}. For $\hat{P}_0^A = \mathbb{I}/\sqrt{2}$, $\hat{P}_1^A = \hat{X}/\sqrt{2}$, $\hat{P}_2^A = \hat{Y}/\sqrt{2}$, and $\hat{P}_3^A = \hat{Z}/\sqrt{2}$ we have
\begin{align}
  &\omega_j^0 = w_1\left(C^j_{Z} - C^j_{-Z}\right), \\
  &\omega_j^1 = w_2\left(C^j_{X} - C^j_{-X}\right), \\
  &\omega_j^2 = -w_2\left(C^j_{Y} - C^j_{-Y}\right), \\
  &\omega_j^3 = \frac{1}{2}\left(C^j_{Z} - C^j_{-Z}\right),
\end{align}
where
\begin{align}
  w_1 &=  \frac{1}{2}\bra{\Psi_{I}}_{AA'}\iden\otimes\hat{Z}\ket{\Psi_{I}}_{AA'} = \frac{1}{2}\left(2b-1\right),\\
  \nonumber w_2 &= \frac{1}{2}\bra{\Psi_{I}}_{AA'}\hat{X}\otimes\hat{X}\ket{\Psi_{I}}_{AA'} \\
  &= -\frac{1}{2}\bra{\Psi_{I}}_{AA'}\hat{Y}\otimes\hat{Y}\ket{\Psi_{I}}_{AA'} = \sqrt{b(1-b)}.
\end{align}
There is one constraint on Alice's reduced state, given by
\begin{align}
  \omega_0^3 = \frac{1}{2}\bra{\Psi_{I}}_{AA'}\hat{Z}\otimes\iden\ket{\Psi_{I}}_{AA'} = w_1.
\end{align}
The full expression for $\rho_{AB}$ is
\begin{align}
  \nonumber\rho_{AB} &= \frac{1}{4}\Big(\mathbb{I}\otimes\mathbb{I} + \omega_0^3\hat{Z}\otimes\mathbb{I} + \omega^0_1\iden\otimes\hat{X} + \omega^0_2\iden\otimes\hat{Y} \\
  \nonumber&+ \omega^0_3\iden\otimes\hat{Z} + \omega^1_1\hat{X}\otimes\hat{X} + \omega^1_2\hat{X}\otimes\hat{Y} + \omega^1_3\hat{X}\otimes\hat{Z} \\
  \nonumber&+ \omega^2_1\hat{Y}\otimes\hat{X} + \omega^2_2\hat{Y}\otimes\hat{Y} + \omega^2_3\hat{Y}\otimes\hat{Z} + \omega^3_1\hat{Z}\otimes\hat{X} \\
  &+ \omega^3_2\hat{Z}\otimes\hat{Y} + \omega^3_3\hat{Z}\otimes\hat{Z}\Big), \label{eqn:rhoAB_C}
\end{align}
where each $\omega^k_j$ contains two free parameters $C_\mu^j$, except for $\omega_0^3$ which is fixed, as well as the pairs $\omega^0_j$ and $\omega^3_j$ which both involve $C_Z^j$. In this calculation we have made no distinction between $\hat{\Omega}_j^k$ and $\hat{\Gamma}_j^k$, and impose the constraints from the known expectation values of the $\hat{\Gamma}_j^k$ as constraints on $\rho_{AB}$ in our convex optimization over the parameters $C^j_\mu$ via expressions of the form
\begin{align}
  {\rm Tr}\left[\rho_{AB}\hat{\Gamma}_j^k\right] = \gamma_j^k.
\end{align}
Note that we have pulled out a factor of $1/2$ from $\omega^k_j$ in Eq.~\eqref{eqn:rhoAB_C}.

\section{Numerical Calculation of the Secret Key Rate}
\label{app:numcalc}

\subsection{Problem Description and Dimensional Reduction}

We follow the general QKD scheme described in \cite{Winick:2018aa}, using a source replacement scheme to describe the prepare-and-measure protocols under consideration. Alice sends one half of an initial state $\ket{\Psi^I_{AB}}$ to Bob, and Eve has access to this half of the state before it reaches Bob. Alice and Bob perform measurements on their halves of the final state, described by a set of POVMs, and record the basis they chose to measure in the Hilbert spaces $\widetilde{A}$ and $\widetilde{B}$. They record the outcomes of their measurements in Hilbert spaces $\overline{A}$ and $\overline{B}$. For simplicity, we assume that there is a one-to-one mapping of outcomes to key-characters, so that the result of the key-map is stored in the register $\overline{A}$ directly, without the need to introduce another register to store the key-map results (as would be in the most general protocol described in \cite{Winick:2018aa}).

Focussing on the specific example of BB84, the measurement and key-map POVM is described by the operators
\begin{align}
  \nonumber\hat{K}_{Z}^A &= \sqrt{p}\Big(\ket{0}^{\widetilde{A}}\otimes\iden^{\widetilde{B}}\otimes\ket{0}^{\overline{A}}\otimes\iden^{\overline{B}}\otimes\ketbra{0}^A\otimes\iden^B \\ &+ \ket{0}^{\widetilde{A}}\otimes\iden^{\widetilde{B}}\otimes\ket{1}^{\overline{A}}\otimes\iden^{\overline{B}}\otimes\ketbra{1}^A\otimes\iden^B\Big), \\
  \nonumber\hat{K}_{X}^A &= \sqrt{1-p}\Big(\ket{1}^{\widetilde{A}}\otimes\iden^{\widetilde{B}}\otimes\ket{0}^{\overline{A}}\otimes\iden^{\overline{B}}\otimes\ketbra{+}^A\otimes\iden^B \\ &+ \ket{1}^{\widetilde{A}}\otimes\iden^{\widetilde{B}}\otimes\ket{1}^{\overline{A}}\otimes\iden^{\overline{B}}\otimes\ketbra{-}^A\otimes\iden^B\Big), \\
  \nonumber\hat{K}_{Z}^B &=  \sqrt{p}\Big(\iden^{\widetilde{A}}\otimes\ket{0}^{\widetilde{B}}\otimes\iden^{\overline{A}}\otimes\ket{0}^{\overline{B}}\otimes\iden^A\otimes\ketbra{0}^B \\ &+ \iden^{\widetilde{A}}\otimes\ket{0}^{\widetilde{B}}\otimes\iden^{\overline{A}}\otimes\ket{1}^{\overline{B}}\otimes\iden^A\otimes\ketbra{1}^B\Big), \\
  \nonumber\hat{K}_{X}^B &=  \sqrt{1-p}\Big(\iden^{\widetilde{A}}\otimes\ket{1}^{\widetilde{B}}\otimes\iden^{\overline{A}}\otimes\ket{0}^{\overline{B}}\otimes\iden^A\otimes\ketbra{+}^B \\ &+ \iden^{\widetilde{A}}\otimes\ket{1}^{\widetilde{B}}\otimes\iden^{\overline{A}}\otimes\ket{1}^{\overline{B}}\otimes\iden^A\otimes\ketbra{-}^B\Big),
\end{align}
with $p$ the probability that Alice (Bob) measures in the Z-basis. Note that for $\overline{A}$ we use a key-map where the measurement outcome pairs $\{0,+\}$ and $\{1,-\}$ correspond to the same character in the secret key. After these measurements, Alice and Bob will post-select on having measured in the same basis, described by the operator
\begin{align}
  \hat{\Pi} = \ketbra{0}^{\widetilde{A}}\otimes\ketbra{0}^{\widetilde{B}}\otimes\iden_4 + \ketbra{1}^{\widetilde{A}}\otimes\ketbra{1}^{\widetilde{B}}\otimes\iden_4
\end{align}
where $\iden_4$ is a 4-qubit identity operator.

The full operation Alice and Bob apply to the state $\rho_{AB}$ is therefore described by the process
\begin{align}
  \nonumber&\mathcal{G}(\rho_{AB}) = \hat{\Pi}\left(\sum_{a,b\in(Z,X)}\left(\hat{K}^A_a\hat{K}^B_b\right)\rho_{AB}\left(\hat{K}^A_a\hat{K}^B_b\right)^\dagger\right)\hat{\Pi}^\dagger \\
  \nonumber&= \left(\hat{K}^A_Z\hat{K}^B_Z\right)\rho_{AB}\left(\hat{K}^A_Z\hat{K}^B_Z\right)^\dagger + \left(\hat{K}^A_X\hat{K}^B_X\right)\rho_{AB}\left(\hat{K}^A_X\hat{K}^B_X\right)^\dagger \\
  &= p_z^2\ketbra{0}^{\widetilde{A}}\otimes\ketbra{0}^{\widetilde{B}}\otimes\rho_z +p_x^2\ketbra{1}^{\widetilde{A}}\otimes\ketbra{1}^{\widetilde{B}}\otimes\rho_x, \label{eqn:G1}
\end{align}
where we have introduced states $\rho_z$ and $\rho_x$ in the $\overline{A}\overline{B}AB$ Hilbert space defined by
\begin{align}
  &\ketbra{0}^{\widetilde{A}}\otimes\ketbra{0}^{\widetilde{B}}\otimes\rho_z = \frac{1}{p_z^2}\left(\hat{K}^A_Z\hat{K}^B_Z\right)\rho_{AB}\left(\hat{K}^A_Z\hat{K}^B_Z\right)^\dagger, \label{eqn:rhoz}\\
  &\ketbra{1}^{\widetilde{A}}\otimes\ketbra{1}^{\widetilde{B}}\otimes\rho_x = \frac{1}{p_x^2}\left(\hat{K}^A_X\hat{K}^B_X\right)\rho_{AB}\left(\hat{K}^A_X\hat{K}^B_X\right)^\dagger, \label{eqn:rhox}
\end{align}
with $p_z = p$ and $p_x = 1-p$ introduced for brevity. We consider a trivial key-map from $\overline{A}\rightarrow R$ that copies the values of the register $\overline{A}$ and stores them in the register $R$. As such, we can combine the registers $\overline{A}$ and $R$, and act upon $\overline{A}$ as we would $R$ for the rest of the calculation \cite{Lin:2019aa}.

The main numerical task to evaluate the secret key rate is to solve the minimization problem
\begin{align}
  \min_{\rho_{AB}}~D(\mathcal{G}(\rho_{AB})||\mathcal{Z}^{\overline{A}}(\mathcal{G}(\rho_{AB})))
\end{align}
where, as a reminder, $\mathcal{Z}^{\mathcal{H}}$ is a pinching channel acting on Hilbert space $\mathcal{H}$. In the form currently written, while the minimization is only over $2^2 \times 2^2$ Hermitian matrices, the actual operations themselves involve $2^6 \times 2^6$ size matrices. This is far from ideal in terms of both memory usage and computational time, though not a problem for the original method developed in \cite{Winick:2018aa}. However, for the approach used here, the convex approximation to the matrix logarithm creates even larger objects in memory \cite{Fawzi:2018aa}, and we have found that the memory usage was far more than what is commonly available on a single work station.

However, we will now show that
\begin{align}
  \nonumber&D(\mathcal{G}(\rho_{AB})||\mathcal{Z}^{\overline{A}}(\mathcal{G}(\rho_{AB}))) \\
  &= p_z^2D(\rho_{AB}||\mathcal{Z}_Z^A(\rho_{AB})) + p_x^2D(\rho_{AB}||\mathcal{Z}_X^A(\rho_{AB})), \label{eqn:newmin}
\end{align}
where
\begin{align}
  \nonumber\mathcal{Z}_Z^A(\rho_{AB}) &= \left(\ketbra{0}^A\otimes\iden\right)\rho_{AB} \left(\ketbra{0}^A\otimes\iden\right) \\ &+ \left(\ketbra{1}^A\otimes\iden\right)\rho_{AB} \left(\ketbra{1}^A\otimes\iden\right) \\
  \nonumber\mathcal{Z}_X^A(\rho_{AB}) &= \left(\ketbra{+}^A\otimes\iden\right)\rho_{AB} \left(\ketbra{+}^A\otimes\iden\right) \\ &+ \left(\ketbra{-}^A\otimes\iden\right)\rho_{AB} \left(\ketbra{-}^A\otimes\iden\right),
\end{align}
with the subscripts of the pinching channels indicating the basis used. Thus, we only need to deal with $4\times4$ matrices in the optimization, and the secret key rate can be calculated by simultaneous minimization of two relative entropies, with all operations performed in only the $AB$ Hilbert space.

The first step is to prove that
\begin{align}
  \nonumber&D(\mathcal{G}(\rho_{AB})||\mathcal{Z}^{\overline{A}}(\mathcal{G}(\rho_{AB}))) \\&= p_z^2D(\rho_z||\mathcal{Z}^{\overline{A}}(\rho_z)) + p_x^2D(\rho_x||\mathcal{Z}^{\overline{A}}(\rho_x)). \label{eqn:G2}
\end{align}
Looking at the structure of Eq.~\eqref{eqn:G1}, we see that it can be written in block-diagonal form as
\begin{align}
  \mathcal{G}(\rho_{AB}) &= \left(\begin{array}{cccc}
  p^2\rho_z & & & \\
   & 0 & & \\
   & & 0 & \\
   & & & (1-p)^2\rho_x
    \end{array}\right).
\end{align}
With this in mind, we will prove the following lemma to prove Eq.~\eqref{eqn:G2}.

\begin{lemma}
  Given two matrices $M = S \oplus T$ and $M' = S' \oplus T'$, where $\dim(S) = \dim(S')$ and $\dim(T) = \dim(T')$, then
  \begin{align}
  &{\rm Tr}\left[M\log(M)\right] = {\rm Tr}\left[S\log(S)\right] + {\rm Tr}\left[T\log(T)\right], \label{eqn:T1} \\
  &{\rm Tr}\left[M\log(M')\right] = {\rm Tr}\left[S\log(S')\right] + {\rm Tr}\left[T\log(T')\right]. \label{eqn:T2}
  \end{align}
  \label{thr:DP}
\end{lemma}
Using lemma \ref{thr:DP} with the identification $M = \mathcal{G}(\rho_{AB})$ and $M' = \mathcal{Z}^{\overline{A}}(\mathcal{G}(\rho_{AB}))$ proves Eq.~\eqref{eqn:G2}. This identification is possible since the pinching channel does not change the block diagonal structure of $\mathcal{G}(\rho_{AB})$.

We then use the following lemma to prove the full result.
\begin{lemma}
  Given $\rho_\mu$, $\rho_{AB}$, and $\mathcal{Z}_\mu^{\mathcal{H}}$ defined as before, then
  \begin{align}
    &D(\rho_z||\mathcal{Z}^{\overline{A}}(\rho_z)) = D(\rho_{AB}||\mathcal{Z}_Z^A(\rho_{AB})) \\
    &D(\rho_x||\mathcal{Z}^{\overline{A}}(\rho_x)) = D(\rho_{AB}||\mathcal{Z}_X^A(\rho_{AB}))
  \end{align}
  \label{thr:sift}
\end{lemma}
Eq.~\eqref{eqn:newmin} follows from application of this lemma to Eq.~\eqref{eqn:G2}.

Therefore, solving the minimization problem
\begin{align}
  \min_{\rho_{AB}}~p^2D(\rho_{AB}||\mathcal{Z}_Z^A(\rho_{AB})) + (1-p)^2D(\rho_{AB}||\mathcal{Z}_X^A(\rho_{AB})),
\end{align}
is equivalent to solving the original minimization problem, but only involves the original Hilbert space of Alice and Bob. This considerably reduces the memory and computational power required to numerically calculate the secret key rate. We note that applications of lemma \ref{thr:DP} beyond those discussed here likely exist, and that versions of lemma \ref{thr:sift} exits for other QKD protocols \cite{Lin:2019aa}.

\subsubsection{Proof of Lemma \ref{thr:DP}}

This result follows straightforwardly from the fact that
\begin{align}
  \log(S\oplus T) = \log(S)\oplus\log(T),
\end{align}
provided that the eigenvalues of $S$ and $T$ fall within the open set $\left(0,\infty\right)$. However, the quantum relative entropy is defined on the half-closed set $\left[0,\infty\right)$, as we take the convention $0\times\log(0) = 0$. This ensures that $D(\rho||\sigma)$ is finite even when either $\rho$ or $\sigma$ have eigenvalues that are zero, provided that the support of $\rho$ does not intersect the kernel of $\sigma$. Assuming ${\rm supp}(\rho) \cap {\rm ker}(\sigma) = 0$, then our convention ensures that lemma \ref{thr:DP} is valid on $\left[0,\infty\right)$ as desired.

\subsubsection{Proof of Lemma \ref{thr:sift}}

From Eqs.~\eqref{eqn:rhoz}, we have that
\begin{align}
  \rho_z = \hat{L}_Z\rho_{AB}\hat{L}_Z^\dagger
\end{align}
where
\begin{align}
  &\nonumber\hat{L}_Z = \ket{00}^{\overline{A}\overline{B}}\otimes\ketbra{00}{00}^{AB} + \ket{01}^{\overline{A}\overline{B}}\otimes\ketbra{01}{01}^{AB} \\&+ \ket{10}^{\overline{A}\overline{B}}\otimes\ketbra{10}{10}^{AB} + \ket{11}^{\overline{A}\overline{B}}\otimes\ketbra{11}{11}^{AB}.
\end{align}
We have that
\begin{align}
  &\left(\ketbra{0}^{\overline{A}}\otimes\iden_3\right)\hat{L}_Z \\
  \nonumber&= \ket{00}^{\overline{A}\overline{B}}\otimes\ketbra{00}{00}^{AB} + \ket{01}^{\overline{A}\overline{B}}\otimes\ketbra{01}{01}^{AB} \\
  \nonumber&= \left(\iden_2\otimes\ketbra{0}^A\otimes\iden\right)\hat{L}_Z = \hat{L}_Z\left(\iden_2\otimes\ketbra{0}^A\otimes\iden\right),
\end{align}
and
\begin{align}
  &\left(\ketbra{1}^{\overline{A}}\otimes\iden_3\right)\hat{L}_Z \\
  \nonumber&= \ket{10}^{\overline{A}\overline{B}}\otimes\ketbra{10}{10}^{AB} + \ket{11}^{\overline{A}\overline{B}}\otimes\ketbra{11}{11}^{AB} \\
  \nonumber&= \left(\iden_2\otimes\ketbra{1}^A\otimes\iden\right)\hat{L}_Z = \hat{L}_Z\left(\iden_2\otimes\ketbra{1}^A\otimes\iden\right),
\end{align}
and therefore
\begin{align}
  &\nonumber\mathcal{Z}^{\overline{A}}(\rho_z) = \sum_{i=0}^1\left(\ketbra{i}^{\overline{A}}\otimes\iden_3\right)\hat{L}_Z\rho_{AB}\hat{L}_Z^\dagger\left(\ketbra{i}^{\overline{A}}\otimes\iden_3\right) \\
  \nonumber&= \sum_{i=0}^1\left(\iden_2\otimes\ketbra{i}^A\otimes\iden\right)\hat{L}_Z\rho_{AB}\hat{L}_Z^\dagger\left(\iden_2\otimes\ketbra{i}^A\otimes\iden\right) \\
  &= \mathcal{Z}_Z^{A}(\rho_z) = \hat{L}_Z\left(\mathcal{Z}_Z^{A}\left(\rho_{AB}\right)\right)\hat{L}_Z^\dagger.
\end{align}

Thus, we have reduced the problem to showing that
\begin{align}
  D(\hat{L}_Z\rho_{AB}\hat{L}_Z^\dagger||\hat{L}_Z\left(\mathcal{Z}_Z^{A}\left(\rho_{AB}\right)\right)\hat{L}_Z^\dagger) = D(\rho_{AB}||\mathcal{Z}_Z^A(\rho_{AB})).
\end{align}
Using the fact that
\begin{align}
  {L}_Z^\dagger{L}_Z = \iden_2,
\end{align}
we see that
\begin{align}
  &{\rm Tr}\left[\hat{L}_Z\rho_{AB}\hat{L}_Z^\dagger\log(\hat{L}_Z\rho_{AB}\hat{L}_Z^\dagger)\right] \\
  \nonumber&= {\rm Tr}\left[\hat{L}_Z\rho_{AB}\hat{L}_Z^\dagger\hat{L}_Z\log(\rho_{AB})\hat{L}_Z^\dagger\right] = {\rm Tr}\left[\rho_{AB}\log(\rho_{AB})\right],
\end{align}
and
\begin{align}
  &{\rm Tr}\left[\hat{L}_Z\rho_{AB}\hat{L}_Z^\dagger\log(\hat{L}_Z\mathcal{Z}_Z^{A}\left(\rho_{AB}\right)\hat{L}_Z^\dagger)\right] \\
  \nonumber&= {\rm Tr}\left[\hat{L}_Z\rho_{AB}\hat{L}_Z^\dagger\hat{L}_Z\log(\mathcal{Z}_Z^{A}\left(\rho_{AB}\right))\hat{L}_Z^\dagger\right] \\ \nonumber&= {\rm Tr}\left[\rho_{AB}\log(\mathcal{Z}_Z^{A}\left(\rho_{AB}\right))\right],
\end{align}
which completes the proof for $\rho_z$. Starting with
\begin{align}
  &\rho_x = \hat{L}_X\rho_{AB}\hat{L}_X^\dagger, \\
  \nonumber&\hat{L}_X = \ket{00}^{\overline{A}\overline{B}}\otimes\ketbra{++}{++}^{AB} + \ket{01}^{\overline{A}\overline{B}}\otimes\ketbra{+-}{+-}^{AB}\\
   &+ \ket{10}^{\overline{A}\overline{B}}\otimes\ketbra{-+}{-+}^{AB} + \ket{11}^{\overline{A}\overline{B}}\otimes\ketbra{--}{--}^{AB},
\end{align}
and following a similar procedure to that outlined for $\rho_z$ proves the full lemma.

\section{Channel Estimation and Error Correction}
\label{app:EC}

In order to compare the performance of different protocols under these two scenarios, we simulate observed statistics using the depolarizing channel. To put our parameterization of the depolarizing channel in context, for symmetric BB84 ($b=1/2$), the commonly quoted quantum bit-error rate (QBER) is half what we define as the depolarizing probability (channel error), $\epsilon$. We use the depolarizing probability rather than the QBER as for asymmetric protocols there is an inherent bit-error rate even when the channel is perfect ($\epsilon=0$). For a fair comparison we compare protocols with the same error added by the channel, i.e.~the depolarizing probability, rather than the same total error.

As a function of the depolarizing probability the constraints imposed on the SDP by parameter estimation are given by
\begin{align}
  &\gamma_{z/x} = {\rm Tr}\left[\qp_{\rm dep}\left(\rho_{AA'}\right)\hat{E}_{Z/X}\right], \\
  &\gamma_{ij} = {\rm Tr}\left[\qp_{\rm dep}\left(\rho_{AA'}\right)\hat{E}_{ij}\right],
\end{align}
depending on the channel reconciliation protocol chosen. For $b=1/2$ (i.e.~symmetric BB84) the fine-grained constraints reduce to $\gamma_{z/x} = \epsilon/2$ as expected.

From the channel model, we obtain the simulated probability distribution $P(x,y)$ that Alice and Bob would observe from an experiment. In this case, if we assume the error correction can be performed at the Shannon limit, then $leak_{\rm EC} = H(X|Y)$, where $X$ is Alice's raw key and $Y$ is Bob's measurement outcomes. $H(X|Y)$ is the conditional Shannon entropy
\begin{align}
  H(X|Y) = -\sum_{x,y} P(x,y)\log_2P(x|y),
\end{align}
for a joint probability distribution $P(x,y)$ and conditional probability distribution $P(x|y)$. For our purposes, $P(x,y)$ can be calculated via the trace of $\rho_{AB}$ with the appropriate POVM operators for measurement in each basis (see appendix \ref{app:numcalc} for these operators), and is a $4\times4$ matrix of probabilities. We also incorporate our post-selection/sifting into $P(x,y)$, by setting $P(x,y) = 0$ where Alice and Bob did not measure in the same basis (as these events are not used in the key generation), and renormalize $P(x,y)$ by the probability that post selection is successful
\begin{align}
  p_{\rm pass} = p^2+(1-p)^2,
\end{align}
which is the probability that Alice and Bob both measure in \emph{either} the Z- or X-basis. Finally, due to Alice's key-map, we add together the elements of $P(x,y)$ for each pair of Alice's raw measurement outcomes that correspond to the same key character, resulting in a $2\times4$ matrix of probabilities.

\section{Security of the Key Rate Lower Bound}
\label{app:sec}

\begin{figure}[t]
  \includegraphics[width = 0.75\columnwidth]{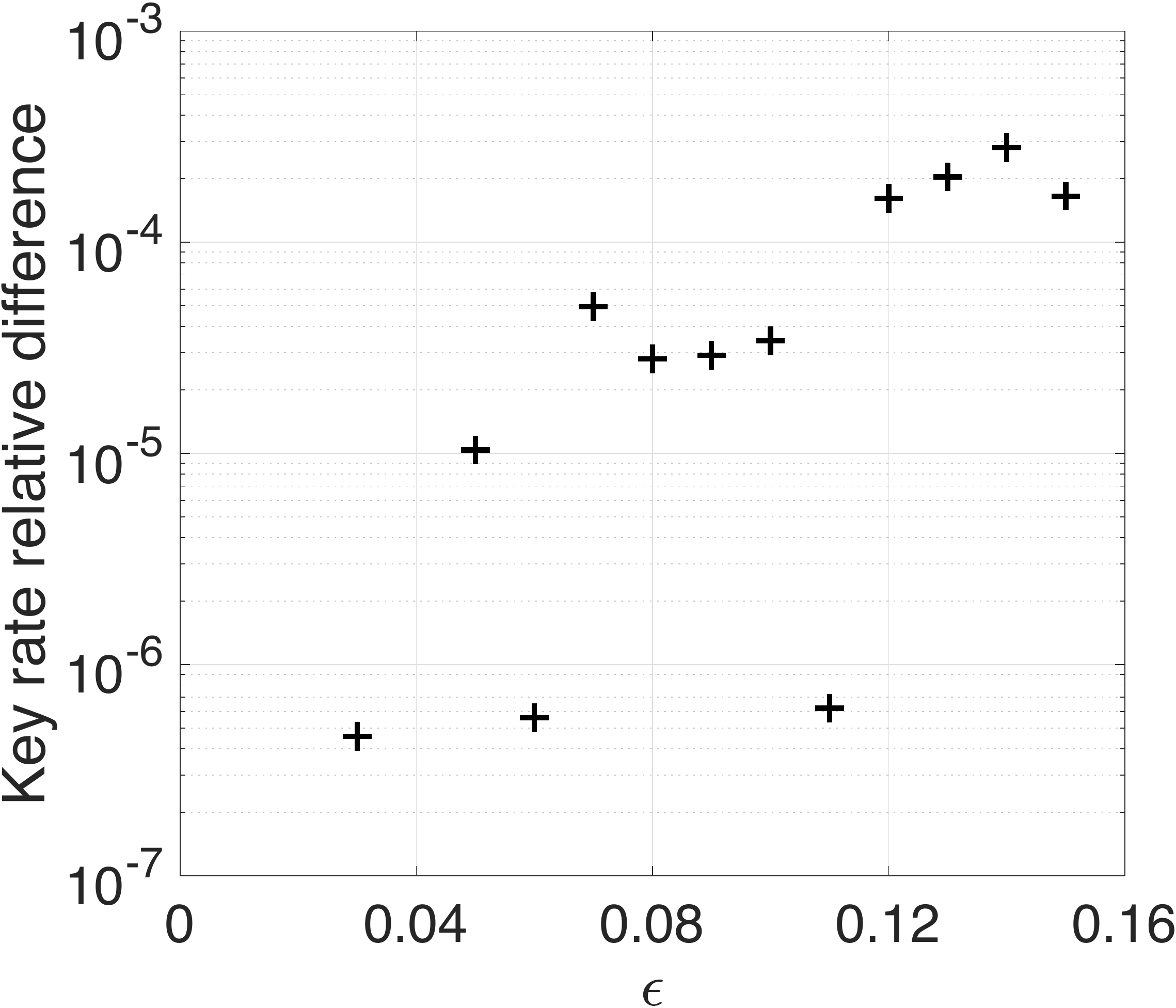}
  \caption{Relative difference between the key rate lower bound calculated using the convex approximation of Ref.~\cite{Fawzi:2018aa}, and that using the dual-linearization of Ref.~\cite{Winick:2018aa}. In both cases the unrestricted symmetric ($b=1/2$) BB84 protocol is considered.}
  \label{fig:appx_dual}
\end{figure}

The numerical approach used in this work, built off the convex approximation of the quantum relative entropy from of Ref.~\cite{Fawzi:2018aa}, solves and approximate version of the dual problem to obtain a lower bound on the secret key rate. To obtain a \emph{secure} lower bound, one can use the techniques of Ref.~\cite{Fawzi:2018aa} to obtain error bounds on the approximation to the quantum relative entropy, in terms of the error bounds on the matrix logarithm. Taking the lower bound of these error bounds gives a secure key rate lower bound. As the order of the approximations used in Ref.~\cite{Fawzi:2018aa} is increased, the calculated secret key rate approaches the actual value, and the error bounds become tighter. For all numerical results presented, we have tested increasing orders of approximation, until the value of the key rate was constant up to a threshold value (typically less than $10^{-4}$). Thus, we are confident that our numerical results for the key rate lower bound are highly accurate.

As further evidence, for the unrestricted symmetric ($b=1/2$) BB84 protocol, we have compared the key rate calculated from our approach (using the approximation of Ref.~\cite{Fawzi:2018aa}) to that calculated using the dual-linearization approach of \cite{Winick:2018aa}. The relative difference between the results of the two approaches is shown in Fig.~\ref{fig:appx_dual}. As can be seen, the key rates are less than a relative difference of $10^{-3}$ away from one another (i.e.~less than a $0.1\%$ difference). We have not applied the second step method of Ref.~\cite{Winick:2018aa} to the restricted situation, as this would require deriving a new linearized dual to the SDP of the restricted situation.

\bibliography{RE_paper.bib}

\begin{thebibliography}{40}%
\makeatletter
\providecommand \@ifxundefined [1]{%
 \@ifx{#1\undefined}
}%
\providecommand \@ifnum [1]{%
 \ifnum #1\expandafter \@firstoftwo
 \else \expandafter \@secondoftwo
 \fi
}%
\providecommand \@ifx [1]{%
 \ifx #1\expandafter \@firstoftwo
 \else \expandafter \@secondoftwo
 \fi
}%
\providecommand \natexlab [1]{#1}%
\providecommand \enquote  [1]{``#1''}%
\providecommand \bibnamefont  [1]{#1}%
\providecommand \bibfnamefont [1]{#1}%
\providecommand \citenamefont [1]{#1}%
\providecommand \href@noop [0]{\@secondoftwo}%
\providecommand \href [0]{\begingroup \@sanitize@url \@href}%
\providecommand \@href[1]{\@@startlink{#1}\@@href}%
\providecommand \@@href[1]{\endgroup#1\@@endlink}%
\providecommand \@sanitize@url [0]{\catcode `\\12\catcode `\$12\catcode
  `\&12\catcode `\#12\catcode `\^12\catcode `\_12\catcode `\%12\relax}%
\providecommand \@@startlink[1]{}%
\providecommand \@@endlink[0]{}%
\providecommand \url  [0]{\begingroup\@sanitize@url \@url }%
\providecommand \@url [1]{\endgroup\@href {#1}{\urlprefix }}%
\providecommand \urlprefix  [0]{URL }%
\providecommand \Eprint [0]{\href }%
\providecommand \doibase [0]{http://dx.doi.org/}%
\providecommand \selectlanguage [0]{\@gobble}%
\providecommand \bibinfo  [0]{\@secondoftwo}%
\providecommand \bibfield  [0]{\@secondoftwo}%
\providecommand \translation [1]{[#1]}%
\providecommand \BibitemOpen [0]{}%
\providecommand \bibitemStop [0]{}%
\providecommand \bibitemNoStop [0]{.\EOS\space}%
\providecommand \EOS [0]{\spacefactor3000\relax}%
\providecommand \BibitemShut  [1]{\csname bibitem#1\endcsname}%
\let\auto@bib@innerbib\@empty
\bibitem [{\citenamefont {{Shor}}(1994)}]{Shor:1994aa}%
  \BibitemOpen
  \bibfield  {author} {\bibinfo {author} {\bibfnamefont {P.~W.}\ \bibnamefont
  {{Shor}}},\ }in\ \href {\doibase 10.1109/SFCS.1994.365700} {\emph {\bibinfo
  {booktitle} {Proceedings 35th Annual Symposium on Foundations of Computer
  Science}}}\ (\bibinfo {year} {1994})\ pp.\ \bibinfo {pages}
  {124--134}\BibitemShut {NoStop}%
\bibitem [{\citenamefont {Rivest}\ \emph {et~al.}(1978)\citenamefont {Rivest},
  \citenamefont {Shamir},\ and\ \citenamefont {Adleman}}]{Rivest:1978aa}%
  \BibitemOpen
  \bibfield  {author} {\bibinfo {author} {\bibfnamefont {R.~L.}\ \bibnamefont
  {Rivest}}, \bibinfo {author} {\bibfnamefont {A.}~\bibnamefont {Shamir}}, \
  and\ \bibinfo {author} {\bibfnamefont {L.}~\bibnamefont {Adleman}},\ }\href
  {\doibase 10.1145/359340.359342} {\bibfield  {journal} {\bibinfo  {journal}
  {Commun. ACM}\ }\textbf {\bibinfo {volume} {21}},\ \bibinfo {pages} {120}
  (\bibinfo {year} {1978})}\BibitemShut {NoStop}%
\bibitem [{\citenamefont {Koblitz}(1987)}]{Koblitz1987}%
  \BibitemOpen
  \bibfield  {author} {\bibinfo {author} {\bibfnamefont {N.}~\bibnamefont
  {Koblitz}},\ }\href@noop {} {\bibfield  {journal} {\bibinfo  {journal}
  {Mathematics of Computation}\ }\textbf {\bibinfo {volume} {48}},\ \bibinfo
  {pages} {203} (\bibinfo {year} {1987})}\BibitemShut {NoStop}%
\bibitem [{\citenamefont {{Diffie}}\ and\ \citenamefont
  {{Hellman}}(1976)}]{Diffie:1976aa}%
  \BibitemOpen
  \bibfield  {author} {\bibinfo {author} {\bibfnamefont {W.}~\bibnamefont
  {{Diffie}}}\ and\ \bibinfo {author} {\bibfnamefont {M.}~\bibnamefont
  {{Hellman}}},\ }\href {\doibase 10.1109/TIT.1976.1055638} {\bibfield
  {journal} {\bibinfo  {journal} {IEEE Transactions on Information Theory}\
  }\textbf {\bibinfo {volume} {22}},\ \bibinfo {pages} {644} (\bibinfo {year}
  {1976})}\BibitemShut {NoStop}%
\bibitem [{\citenamefont {McEliece}(1978)}]{McEliece1978}%
  \BibitemOpen
  \bibfield  {author} {\bibinfo {author} {\bibfnamefont {R.~J.}\ \bibnamefont
  {McEliece}},\ }\href@noop {} {\bibfield  {journal} {\bibinfo  {journal} {Deep
  Space Network Progress Report}\ }\textbf {\bibinfo {volume} {44}},\ \bibinfo
  {pages} {114} (\bibinfo {year} {1978})}\BibitemShut {NoStop}%
\bibitem [{\citenamefont {Hoffstein}\ \emph {et~al.}(1998)\citenamefont
  {Hoffstein}, \citenamefont {Pipher},\ and\ \citenamefont
  {Silverman}}]{Hoffstein:1998}%
  \BibitemOpen
  \bibfield  {author} {\bibinfo {author} {\bibfnamefont {J.}~\bibnamefont
  {Hoffstein}}, \bibinfo {author} {\bibfnamefont {J.}~\bibnamefont {Pipher}}, \
  and\ \bibinfo {author} {\bibfnamefont {J.~H.}\ \bibnamefont {Silverman}},\
  }in\ \href {http://dl.acm.org/citation.cfm?id=648184.749737} {\emph {\bibinfo
  {booktitle} {Proceedings of the Third International Symposium on Algorithmic
  Number Theory}}},\ \bibinfo {series and number} {ANTS-III}\ (\bibinfo
  {publisher} {Springer-Verlag},\ \bibinfo {address} {Berlin, Heidelberg},\
  \bibinfo {year} {1998})\ pp.\ \bibinfo {pages} {267--288}\BibitemShut
  {NoStop}%
\bibitem [{\citenamefont {De~Feo}\ \emph {et~al.}(2014)\citenamefont {De~Feo},
  \citenamefont {Jao},\ and\ \citenamefont {Pl{\^u}t}}]{De-Feo:2014aa}%
  \BibitemOpen
  \bibfield  {author} {\bibinfo {author} {\bibfnamefont {L.}~\bibnamefont
  {De~Feo}}, \bibinfo {author} {\bibfnamefont {D.}~\bibnamefont {Jao}}, \ and\
  \bibinfo {author} {\bibfnamefont {J.}~\bibnamefont {Pl{\^u}t}},\ }\href
  {\doibase 10.1515/jmc-2012-0015} {\bibfield  {journal} {\bibinfo  {journal}
  {Journal of Mathematical Cryptology}\ }\textbf {\bibinfo {volume} {8(3)}},\
  \bibinfo {pages} {209} (\bibinfo {year} {2014})}\BibitemShut {NoStop}%
\bibitem [{\citenamefont {Bernstein}\ and\ \citenamefont
  {Lange}(2017)}]{Bernstein:2017aa}%
  \BibitemOpen
  \bibfield  {author} {\bibinfo {author} {\bibfnamefont {D.~J.}\ \bibnamefont
  {Bernstein}}\ and\ \bibinfo {author} {\bibfnamefont {T.}~\bibnamefont
  {Lange}},\ }\href {https://doi.org/10.1038/nature23461} {\bibfield  {journal}
  {\bibinfo  {journal} {Nature}\ }\textbf {\bibinfo {volume} {549}},\ \bibinfo
  {pages} {188 EP } (\bibinfo {year} {2017})}\BibitemShut {NoStop}%
\bibitem [{\citenamefont {Korzh}\ \emph {et~al.}(2015)\citenamefont {Korzh},
  \citenamefont {Lim}, \citenamefont {Houlmann}, \citenamefont {Gisin},
  \citenamefont {Li}, \citenamefont {Nolan}, \citenamefont {Sanguinetti},
  \citenamefont {Thew},\ and\ \citenamefont {Zbinden}}]{Korzh:2015aa}%
  \BibitemOpen
  \bibfield  {author} {\bibinfo {author} {\bibfnamefont {B.}~\bibnamefont
  {Korzh}}, \bibinfo {author} {\bibfnamefont {C.~C.~W.}\ \bibnamefont {Lim}},
  \bibinfo {author} {\bibfnamefont {R.}~\bibnamefont {Houlmann}}, \bibinfo
  {author} {\bibfnamefont {N.}~\bibnamefont {Gisin}}, \bibinfo {author}
  {\bibfnamefont {M.~J.}\ \bibnamefont {Li}}, \bibinfo {author} {\bibfnamefont
  {D.}~\bibnamefont {Nolan}}, \bibinfo {author} {\bibfnamefont
  {B.}~\bibnamefont {Sanguinetti}}, \bibinfo {author} {\bibfnamefont
  {R.}~\bibnamefont {Thew}}, \ and\ \bibinfo {author} {\bibfnamefont
  {H.}~\bibnamefont {Zbinden}},\ }\href
  {https://doi.org/10.1038/nphoton.2014.327} {\bibfield  {journal} {\bibinfo
  {journal} {Nature Photonics}\ }\textbf {\bibinfo {volume} {9}},\ \bibinfo
  {pages} {163} (\bibinfo {year} {2015})}\BibitemShut {NoStop}%
\bibitem [{\citenamefont {Yin}\ \emph {et~al.}(2016)\citenamefont {Yin},
  \citenamefont {Chen}, \citenamefont {Yu}, \citenamefont {Liu}, \citenamefont
  {You}, \citenamefont {Zhou}, \citenamefont {Chen}, \citenamefont {Mao},
  \citenamefont {Huang}, \citenamefont {Zhang}, \citenamefont {Chen},
  \citenamefont {Li}, \citenamefont {Nolan}, \citenamefont {Zhou},
  \citenamefont {Jiang}, \citenamefont {Wang}, \citenamefont {Zhang},
  \citenamefont {Wang},\ and\ \citenamefont {Pan}}]{Yin:2016aa}%
  \BibitemOpen
  \bibfield  {author} {\bibinfo {author} {\bibfnamefont {H.-L.}\ \bibnamefont
  {Yin}}, \bibinfo {author} {\bibfnamefont {T.-Y.}\ \bibnamefont {Chen}},
  \bibinfo {author} {\bibfnamefont {Z.-W.}\ \bibnamefont {Yu}}, \bibinfo
  {author} {\bibfnamefont {H.}~\bibnamefont {Liu}}, \bibinfo {author}
  {\bibfnamefont {L.-X.}\ \bibnamefont {You}}, \bibinfo {author} {\bibfnamefont
  {Y.-H.}\ \bibnamefont {Zhou}}, \bibinfo {author} {\bibfnamefont {S.-J.}\
  \bibnamefont {Chen}}, \bibinfo {author} {\bibfnamefont {Y.}~\bibnamefont
  {Mao}}, \bibinfo {author} {\bibfnamefont {M.-Q.}\ \bibnamefont {Huang}},
  \bibinfo {author} {\bibfnamefont {W.-J.}\ \bibnamefont {Zhang}}, \bibinfo
  {author} {\bibfnamefont {H.}~\bibnamefont {Chen}}, \bibinfo {author}
  {\bibfnamefont {M.~J.}\ \bibnamefont {Li}}, \bibinfo {author} {\bibfnamefont
  {D.}~\bibnamefont {Nolan}}, \bibinfo {author} {\bibfnamefont
  {F.}~\bibnamefont {Zhou}}, \bibinfo {author} {\bibfnamefont {X.}~\bibnamefont
  {Jiang}}, \bibinfo {author} {\bibfnamefont {Z.}~\bibnamefont {Wang}},
  \bibinfo {author} {\bibfnamefont {Q.}~\bibnamefont {Zhang}}, \bibinfo
  {author} {\bibfnamefont {X.-B.}\ \bibnamefont {Wang}}, \ and\ \bibinfo
  {author} {\bibfnamefont {J.-W.}\ \bibnamefont {Pan}},\ }\href {\doibase
  10.1103/PhysRevLett.117.190501} {\bibfield  {journal} {\bibinfo  {journal}
  {Phys. Rev. Lett.}\ }\textbf {\bibinfo {volume} {117}},\ \bibinfo {pages}
  {190501} (\bibinfo {year} {2016})}\BibitemShut {NoStop}%
\bibitem [{\citenamefont {Bunandar}\ \emph {et~al.}(2018)\citenamefont
  {Bunandar}, \citenamefont {Lentine}, \citenamefont {Lee}, \citenamefont
  {Cai}, \citenamefont {Long}, \citenamefont {Boynton}, \citenamefont
  {Martinez}, \citenamefont {DeRose}, \citenamefont {Chen}, \citenamefont
  {Grein}, \citenamefont {Trotter}, \citenamefont {Starbuck}, \citenamefont
  {Pomerene}, \citenamefont {Hamilton}, \citenamefont {Wong}, \citenamefont
  {Camacho}, \citenamefont {Davids}, \citenamefont {Urayama},\ and\
  \citenamefont {Englund}}]{Bunandar:2018aa}%
  \BibitemOpen
  \bibfield  {author} {\bibinfo {author} {\bibfnamefont {D.}~\bibnamefont
  {Bunandar}}, \bibinfo {author} {\bibfnamefont {A.}~\bibnamefont {Lentine}},
  \bibinfo {author} {\bibfnamefont {C.}~\bibnamefont {Lee}}, \bibinfo {author}
  {\bibfnamefont {H.}~\bibnamefont {Cai}}, \bibinfo {author} {\bibfnamefont
  {C.~M.}\ \bibnamefont {Long}}, \bibinfo {author} {\bibfnamefont
  {N.}~\bibnamefont {Boynton}}, \bibinfo {author} {\bibfnamefont
  {N.}~\bibnamefont {Martinez}}, \bibinfo {author} {\bibfnamefont
  {C.}~\bibnamefont {DeRose}}, \bibinfo {author} {\bibfnamefont
  {C.}~\bibnamefont {Chen}}, \bibinfo {author} {\bibfnamefont {M.}~\bibnamefont
  {Grein}}, \bibinfo {author} {\bibfnamefont {D.}~\bibnamefont {Trotter}},
  \bibinfo {author} {\bibfnamefont {A.}~\bibnamefont {Starbuck}}, \bibinfo
  {author} {\bibfnamefont {A.}~\bibnamefont {Pomerene}}, \bibinfo {author}
  {\bibfnamefont {S.}~\bibnamefont {Hamilton}}, \bibinfo {author}
  {\bibfnamefont {F.~N.~C.}\ \bibnamefont {Wong}}, \bibinfo {author}
  {\bibfnamefont {R.}~\bibnamefont {Camacho}}, \bibinfo {author} {\bibfnamefont
  {P.}~\bibnamefont {Davids}}, \bibinfo {author} {\bibfnamefont
  {J.}~\bibnamefont {Urayama}}, \ and\ \bibinfo {author} {\bibfnamefont
  {D.}~\bibnamefont {Englund}},\ }\href {\doibase 10.1103/PhysRevX.8.021009}
  {\bibfield  {journal} {\bibinfo  {journal} {Phys. Rev. X}\ }\textbf {\bibinfo
  {volume} {8}},\ \bibinfo {pages} {021009} (\bibinfo {year}
  {2018})}\BibitemShut {NoStop}%
\bibitem [{\citenamefont {Pirandola}\ \emph
  {et~al.}(2017{\natexlab{a}})\citenamefont {Pirandola}, \citenamefont
  {Laurenza}, \citenamefont {Ottaviani},\ and\ \citenamefont
  {Banchi}}]{Pirandola:2017cj}%
  \BibitemOpen
  \bibfield  {author} {\bibinfo {author} {\bibfnamefont {S.}~\bibnamefont
  {Pirandola}}, \bibinfo {author} {\bibfnamefont {R.}~\bibnamefont {Laurenza}},
  \bibinfo {author} {\bibfnamefont {C.}~\bibnamefont {Ottaviani}}, \ and\
  \bibinfo {author} {\bibfnamefont {L.}~\bibnamefont {Banchi}},\ }\href
  {\doibase 10.1038/ncomms15043} {\bibfield  {journal} {\bibinfo  {journal}
  {Nature Communications}\ }\textbf {\bibinfo {volume} {8}},\ \bibinfo {pages}
  {15043} (\bibinfo {year} {2017}{\natexlab{a}})}\BibitemShut {NoStop}%
\bibitem [{\citenamefont {Muralidharan}\ \emph {et~al.}(2016)\citenamefont
  {Muralidharan}, \citenamefont {Li}, \citenamefont {Kim}, \citenamefont
  {L{\"u}tkenhaus}, \citenamefont {Lukin},\ and\ \citenamefont
  {Jiang}}]{Muralidharan:2016ye}%
  \BibitemOpen
  \bibfield  {author} {\bibinfo {author} {\bibfnamefont {S.}~\bibnamefont
  {Muralidharan}}, \bibinfo {author} {\bibfnamefont {L.}~\bibnamefont {Li}},
  \bibinfo {author} {\bibfnamefont {J.}~\bibnamefont {Kim}}, \bibinfo {author}
  {\bibfnamefont {N.}~\bibnamefont {L{\"u}tkenhaus}}, \bibinfo {author}
  {\bibfnamefont {M.~D.}\ \bibnamefont {Lukin}}, \ and\ \bibinfo {author}
  {\bibfnamefont {L.}~\bibnamefont {Jiang}},\ }\href
  {https://doi.org/10.1038/srep20463} {\bibfield  {journal} {\bibinfo
  {journal} {Scientific Reports}\ }\textbf {\bibinfo {volume} {6}},\ \bibinfo
  {pages} {20463} (\bibinfo {year} {2016})}\BibitemShut {NoStop}%
\bibitem [{\citenamefont {Krovi}\ \emph {et~al.}(2016)\citenamefont {Krovi},
  \citenamefont {Guha}, \citenamefont {Dutton}, \citenamefont {Slater},
  \citenamefont {Simon},\ and\ \citenamefont {Tittel}}]{Krovi2016}%
  \BibitemOpen
  \bibfield  {author} {\bibinfo {author} {\bibfnamefont {H.}~\bibnamefont
  {Krovi}}, \bibinfo {author} {\bibfnamefont {S.}~\bibnamefont {Guha}},
  \bibinfo {author} {\bibfnamefont {Z.}~\bibnamefont {Dutton}}, \bibinfo
  {author} {\bibfnamefont {J.~A.}\ \bibnamefont {Slater}}, \bibinfo {author}
  {\bibfnamefont {C.}~\bibnamefont {Simon}}, \ and\ \bibinfo {author}
  {\bibfnamefont {W.}~\bibnamefont {Tittel}},\ }\href {\doibase
  10.1007/s00340-015-6297-4} {\bibfield  {journal} {\bibinfo  {journal}
  {Applied Physics B}\ }\textbf {\bibinfo {volume} {122}},\ \bibinfo {pages}
  {52} (\bibinfo {year} {2016})}\BibitemShut {NoStop}%
\bibitem [{\citenamefont {Guha}\ \emph {et~al.}(2015)\citenamefont {Guha},
  \citenamefont {Krovi}, \citenamefont {Fuchs}, \citenamefont {Dutton},
  \citenamefont {Slater}, \citenamefont {Simon},\ and\ \citenamefont
  {Tittel}}]{PhysRevA.92.022357}%
  \BibitemOpen
  \bibfield  {author} {\bibinfo {author} {\bibfnamefont {S.}~\bibnamefont
  {Guha}}, \bibinfo {author} {\bibfnamefont {H.}~\bibnamefont {Krovi}},
  \bibinfo {author} {\bibfnamefont {C.~A.}\ \bibnamefont {Fuchs}}, \bibinfo
  {author} {\bibfnamefont {Z.}~\bibnamefont {Dutton}}, \bibinfo {author}
  {\bibfnamefont {J.~A.}\ \bibnamefont {Slater}}, \bibinfo {author}
  {\bibfnamefont {C.}~\bibnamefont {Simon}}, \ and\ \bibinfo {author}
  {\bibfnamefont {W.}~\bibnamefont {Tittel}},\ }\href {\doibase
  10.1103/PhysRevA.92.022357} {\bibfield  {journal} {\bibinfo  {journal} {Phys.
  Rev. A}\ }\textbf {\bibinfo {volume} {92}},\ \bibinfo {pages} {022357}
  (\bibinfo {year} {2015})}\BibitemShut {NoStop}%
\bibitem [{\citenamefont {Pirandola}\ \emph
  {et~al.}(2017{\natexlab{b}})\citenamefont {Pirandola}, \citenamefont
  {Laurenza}, \citenamefont {Ottaviani},\ and\ \citenamefont {Banchi}}]{PLOB}%
  \BibitemOpen
  \bibfield  {author} {\bibinfo {author} {\bibfnamefont {S.}~\bibnamefont
  {Pirandola}}, \bibinfo {author} {\bibfnamefont {R.}~\bibnamefont {Laurenza}},
  \bibinfo {author} {\bibfnamefont {C.}~\bibnamefont {Ottaviani}}, \ and\
  \bibinfo {author} {\bibfnamefont {L.}~\bibnamefont {Banchi}},\ }\href
  {https://doi.org/10.1038/ncomms15043} {\bibfield  {journal} {\bibinfo
  {journal} {Nature Communications}\ }\textbf {\bibinfo {volume} {8}},\
  \bibinfo {pages} {15043 EP } (\bibinfo {year} {2017}{\natexlab{b}})},\
  \bibinfo {note} {article}\BibitemShut {NoStop}%
\bibitem [{\citenamefont {Takeoka}\ \emph {et~al.}(2014)\citenamefont
  {Takeoka}, \citenamefont {Guha},\ and\ \citenamefont {Wilde}}]{Takeoka2014}%
  \BibitemOpen
  \bibfield  {author} {\bibinfo {author} {\bibfnamefont {M.}~\bibnamefont
  {Takeoka}}, \bibinfo {author} {\bibfnamefont {S.}~\bibnamefont {Guha}}, \
  and\ \bibinfo {author} {\bibfnamefont {M.~M.}\ \bibnamefont {Wilde}},\ }\href
  {https://doi.org/10.1038/ncomms6235} {\bibfield  {journal} {\bibinfo
  {journal} {Nature Communications}\ }\textbf {\bibinfo {volume} {5}},\
  \bibinfo {pages} {5235 EP } (\bibinfo {year} {2014})},\ \bibinfo {note}
  {article}\BibitemShut {NoStop}%
\bibitem [{\citenamefont {Sangouard}\ \emph {et~al.}(2011)\citenamefont
  {Sangouard}, \citenamefont {Simon}, \citenamefont {de~Riedmatten},\ and\
  \citenamefont {Gisin}}]{RevModPhys.83.33}%
  \BibitemOpen
  \bibfield  {author} {\bibinfo {author} {\bibfnamefont {N.}~\bibnamefont
  {Sangouard}}, \bibinfo {author} {\bibfnamefont {C.}~\bibnamefont {Simon}},
  \bibinfo {author} {\bibfnamefont {H.}~\bibnamefont {de~Riedmatten}}, \ and\
  \bibinfo {author} {\bibfnamefont {N.}~\bibnamefont {Gisin}},\ }\href
  {\doibase 10.1103/RevModPhys.83.33} {\bibfield  {journal} {\bibinfo
  {journal} {Rev. Mod. Phys.}\ }\textbf {\bibinfo {volume} {83}},\ \bibinfo
  {pages} {33} (\bibinfo {year} {2011})}\BibitemShut {NoStop}%
\bibitem [{\citenamefont {Preskill}(2018)}]{Preskill:2018}%
  \BibitemOpen
  \bibfield  {author} {\bibinfo {author} {\bibfnamefont {J.}~\bibnamefont
  {Preskill}},\ }\href {\doibase 10.22331/q-2018-08-06-79} {\bibfield
  {journal} {\bibinfo  {journal} {{Quantum}}\ }\textbf {\bibinfo {volume}
  {2}},\ \bibinfo {pages} {79} (\bibinfo {year} {2018})}\BibitemShut {NoStop}%
\bibitem [{\citenamefont {Gottesman}(1999)}]{Gottesman:1999aa}%
  \BibitemOpen
  \bibfield  {author} {\bibinfo {author} {\bibfnamefont {D.}~\bibnamefont
  {Gottesman}},\ }in\ \href@noop {} {\emph {\bibinfo {booktitle} {Group22:
  Proceedings of the XXII International Colloquium on Group Theoretical Methods
  in Physics}}},\ \bibinfo {editor} {edited by\ \bibinfo {editor}
  {\bibfnamefont {S.~P.}\ \bibnamefont {Corney}}, \bibinfo {editor}
  {\bibfnamefont {R.}~\bibnamefont {Delbourgo}}, \ and\ \bibinfo {editor}
  {\bibfnamefont {P.~D.}\ \bibnamefont {Jarvis}}}\ (\bibinfo  {publisher}
  {International Press},\ \bibinfo {address} {Cambridge, MA},\ \bibinfo {year}
  {1999})\ pp.\ \bibinfo {pages} {32--43}\BibitemShut {NoStop}%
\bibitem [{\citenamefont {Bennett}\ and\ \citenamefont
  {Brassard}(1984)}]{Bennett:1984aa}%
  \BibitemOpen
  \bibfield  {author} {\bibinfo {author} {\bibfnamefont {C.}~\bibnamefont
  {Bennett}}\ and\ \bibinfo {author} {\bibfnamefont {G.}~\bibnamefont
  {Brassard}},\ }in\ \href@noop {} {\emph {\bibinfo {booktitle} {Proceedings of
  IEEE International Conference on Computers, Systems, and Signal Processing,
  Bangalore, India}}}\ (\bibinfo  {publisher} {IEEE},\ \bibinfo {address} {New
  York},\ \bibinfo {year} {1984})\BibitemShut {NoStop}%
\bibitem [{\citenamefont {Coles}\ \emph {et~al.}(2016)\citenamefont {Coles},
  \citenamefont {Metodiev},\ and\ \citenamefont
  {L{\"u}tkenhaus}}]{Coles:2016aa}%
  \BibitemOpen
  \bibfield  {author} {\bibinfo {author} {\bibfnamefont {P.~J.}\ \bibnamefont
  {Coles}}, \bibinfo {author} {\bibfnamefont {E.~M.}\ \bibnamefont {Metodiev}},
  \ and\ \bibinfo {author} {\bibfnamefont {N.}~\bibnamefont {L{\"u}tkenhaus}},\
  }\href {https://doi.org/10.1038/ncomms11712} {\bibfield  {journal} {\bibinfo
  {journal} {Nature Communications}\ }\textbf {\bibinfo {volume} {7}},\
  \bibinfo {pages} {11712} (\bibinfo {year} {2016})}\BibitemShut {NoStop}%
\bibitem [{\citenamefont {Winick}\ \emph {et~al.}(2018)\citenamefont {Winick},
  \citenamefont {L{\"{u}}tkenhaus},\ and\ \citenamefont
  {Coles}}]{Winick:2018aa}%
  \BibitemOpen
  \bibfield  {author} {\bibinfo {author} {\bibfnamefont {A.}~\bibnamefont
  {Winick}}, \bibinfo {author} {\bibfnamefont {N.}~\bibnamefont
  {L{\"{u}}tkenhaus}}, \ and\ \bibinfo {author} {\bibfnamefont {P.~J.}\
  \bibnamefont {Coles}},\ }\href {\doibase 10.22331/q-2018-07-26-77} {\bibfield
   {journal} {\bibinfo  {journal} {{Quantum}}\ }\textbf {\bibinfo {volume}
  {2}},\ \bibinfo {pages} {77} (\bibinfo {year} {2018})}\BibitemShut {NoStop}%
\bibitem [{\citenamefont {Damgaard}\ \emph {et~al.}(2007)\citenamefont
  {Damgaard}, \citenamefont {Fehr}, \citenamefont {Renner}, \citenamefont
  {Salvail},\ and\ \citenamefont {Schaffner}}]{Damgaard:fq}%
  \BibitemOpen
  \bibfield  {author} {\bibinfo {author} {\bibfnamefont {I.~B.}\ \bibnamefont
  {Damgaard}}, \bibinfo {author} {\bibfnamefont {S.}~\bibnamefont {Fehr}},
  \bibinfo {author} {\bibfnamefont {R.}~\bibnamefont {Renner}}, \bibinfo
  {author} {\bibfnamefont {L.}~\bibnamefont {Salvail}}, \ and\ \bibinfo
  {author} {\bibfnamefont {C.}~\bibnamefont {Schaffner}},\ }\href@noop {}
  {\bibfield  {journal} {\bibinfo  {journal} {full version of CRYPTO 2007, LNCS
  4622}\ } (\bibinfo {year} {2007})}\BibitemShut {NoStop}%
\bibitem [{\citenamefont {Hosseinidehaj}\ \emph {et~al.}(2019)\citenamefont
  {Hosseinidehaj}, \citenamefont {Walk},\ and\ \citenamefont
  {Ralph}}]{Hosseinidehaj:2019aa}%
  \BibitemOpen
  \bibfield  {author} {\bibinfo {author} {\bibfnamefont {N.}~\bibnamefont
  {Hosseinidehaj}}, \bibinfo {author} {\bibfnamefont {N.}~\bibnamefont {Walk}},
  \ and\ \bibinfo {author} {\bibfnamefont {T.~C.}\ \bibnamefont {Ralph}},\
  }\href {\doibase 10.1103/PhysRevA.99.052336} {\bibfield  {journal} {\bibinfo
  {journal} {Phys. Rev. A}\ }\textbf {\bibinfo {volume} {99}},\ \bibinfo
  {pages} {052336} (\bibinfo {year} {2019})}\BibitemShut {NoStop}%
\bibitem [{\citenamefont {Pan}\ \emph {et~al.}(2019)\citenamefont {Pan},
  \citenamefont {Seshadreesan}, \citenamefont {Clark}, \citenamefont {Adcock},
  \citenamefont {Djordjevic}, \citenamefont {Shapiro},\ and\ \citenamefont
  {Guha}}]{Pan:2019aa}%
  \BibitemOpen
  \bibfield  {author} {\bibinfo {author} {\bibfnamefont {Z.}~\bibnamefont
  {Pan}}, \bibinfo {author} {\bibfnamefont {K.~P.}\ \bibnamefont
  {Seshadreesan}}, \bibinfo {author} {\bibfnamefont {W.}~\bibnamefont {Clark}},
  \bibinfo {author} {\bibfnamefont {M.~R.}\ \bibnamefont {Adcock}}, \bibinfo
  {author} {\bibfnamefont {I.~B.}\ \bibnamefont {Djordjevic}}, \bibinfo
  {author} {\bibfnamefont {J.~H.}\ \bibnamefont {Shapiro}}, \ and\ \bibinfo
  {author} {\bibfnamefont {S.}~\bibnamefont {Guha}},\ }\href@noop {} {\enquote
  {\bibinfo {title} {Secret key distillation across a quantum wiretap channel
  under restricted eavesdropping},}\ } (\bibinfo {year} {2019}),\ \Eprint
  {http://arxiv.org/abs/arXiv:1903.03136} {arXiv:1903.03136} \BibitemShut
  {NoStop}%
\bibitem [{\citenamefont {Ferenczi}\ and\ \citenamefont
  {L\"utkenhaus}(2012)}]{Ferenczi:2012aa}%
  \BibitemOpen
  \bibfield  {author} {\bibinfo {author} {\bibfnamefont {A.}~\bibnamefont
  {Ferenczi}}\ and\ \bibinfo {author} {\bibfnamefont {N.}~\bibnamefont
  {L\"utkenhaus}},\ }\href {\doibase 10.1103/PhysRevA.85.052310} {\bibfield
  {journal} {\bibinfo  {journal} {Phys. Rev. A}\ }\textbf {\bibinfo {volume}
  {85}},\ \bibinfo {pages} {052310} (\bibinfo {year} {2012})}\BibitemShut
  {NoStop}%
\bibitem [{\citenamefont {Bru\ss{}}(1998)}]{Bruss:1998aa}%
  \BibitemOpen
  \bibfield  {author} {\bibinfo {author} {\bibfnamefont {D.}~\bibnamefont
  {Bru\ss{}}},\ }\href {\doibase 10.1103/PhysRevLett.81.3018} {\bibfield
  {journal} {\bibinfo  {journal} {Phys. Rev. Lett.}\ }\textbf {\bibinfo
  {volume} {81}},\ \bibinfo {pages} {3018} (\bibinfo {year}
  {1998})}\BibitemShut {NoStop}%
\bibitem [{\citenamefont {Bennett}\ \emph {et~al.}(1992)\citenamefont
  {Bennett}, \citenamefont {Brassard},\ and\ \citenamefont
  {Mermin}}]{Bennett:1992aa}%
  \BibitemOpen
  \bibfield  {author} {\bibinfo {author} {\bibfnamefont {C.~H.}\ \bibnamefont
  {Bennett}}, \bibinfo {author} {\bibfnamefont {G.}~\bibnamefont {Brassard}}, \
  and\ \bibinfo {author} {\bibfnamefont {N.~D.}\ \bibnamefont {Mermin}},\
  }\href {\doibase 10.1103/PhysRevLett.68.557} {\bibfield  {journal} {\bibinfo
  {journal} {Phys. Rev. Lett.}\ }\textbf {\bibinfo {volume} {68}},\ \bibinfo
  {pages} {557} (\bibinfo {year} {1992})}\BibitemShut {NoStop}%
\bibitem [{\citenamefont {Curty}\ \emph {et~al.}(2004)\citenamefont {Curty},
  \citenamefont {Lewenstein},\ and\ \citenamefont
  {L\"utkenhaus}}]{Curty:2004aa}%
  \BibitemOpen
  \bibfield  {author} {\bibinfo {author} {\bibfnamefont {M.}~\bibnamefont
  {Curty}}, \bibinfo {author} {\bibfnamefont {M.}~\bibnamefont {Lewenstein}}, \
  and\ \bibinfo {author} {\bibfnamefont {N.}~\bibnamefont {L\"utkenhaus}},\
  }\href {\doibase 10.1103/PhysRevLett.92.217903} {\bibfield  {journal}
  {\bibinfo  {journal} {Phys. Rev. Lett.}\ }\textbf {\bibinfo {volume} {92}},\
  \bibinfo {pages} {217903} (\bibinfo {year} {2004})}\BibitemShut {NoStop}%
\bibitem [{\citenamefont {Nielsen}\ and\ \citenamefont
  {Chuang}(2000)}]{Nielsen00}%
  \BibitemOpen
  \bibfield  {author} {\bibinfo {author} {\bibfnamefont {M.}~\bibnamefont
  {Nielsen}}\ and\ \bibinfo {author} {\bibfnamefont {I.}~\bibnamefont
  {Chuang}},\ }\href@noop {} {\emph {\bibinfo {title} {Quantum Computation and
  Quantum Information}}}\ (\bibinfo  {publisher} {Cambridge University Press},\
  \bibinfo {address} {Cambridge, UK},\ \bibinfo {year} {2000})\BibitemShut
  {NoStop}%
\bibitem [{\citenamefont {Govia}\ \emph {et~al.}(2019)\citenamefont {Govia},
  \citenamefont {Ribeill}, \citenamefont {Rist{\`e}}, \citenamefont {Ware},\
  and\ \citenamefont {Krovi}}]{PAPA}%
  \BibitemOpen
  \bibfield  {author} {\bibinfo {author} {\bibfnamefont {L.~C.~G.}\
  \bibnamefont {Govia}}, \bibinfo {author} {\bibfnamefont {G.~J.}\ \bibnamefont
  {Ribeill}}, \bibinfo {author} {\bibfnamefont {D.}~\bibnamefont {Rist{\`e}}},
  \bibinfo {author} {\bibfnamefont {M.}~\bibnamefont {Ware}}, \ and\ \bibinfo
  {author} {\bibfnamefont {H.}~\bibnamefont {Krovi}},\ }\href@noop {} {\enquote
  {\bibinfo {title} {{Bootstrapping quantum process tomography via a
  perturbative ansatz}},}\ } (\bibinfo {year} {2019}),\ \bibinfo {note}
  {arXiv:1902.10821}\BibitemShut {NoStop}%
\bibitem [{\citenamefont {Coles}(2012)}]{Coles:2012aa}%
  \BibitemOpen
  \bibfield  {author} {\bibinfo {author} {\bibfnamefont {P.~J.}\ \bibnamefont
  {Coles}},\ }\href {\doibase 10.1103/PhysRevA.85.042103} {\bibfield  {journal}
  {\bibinfo  {journal} {Phys. Rev. A}\ }\textbf {\bibinfo {volume} {85}},\
  \bibinfo {pages} {042103} (\bibinfo {year} {2012})}\BibitemShut {NoStop}%
\bibitem [{\citenamefont {Fawzi}\ \emph {et~al.}(2018)\citenamefont {Fawzi},
  \citenamefont {Saunderson},\ and\ \citenamefont {Parrilo}}]{Fawzi:2018aa}%
  \BibitemOpen
  \bibfield  {author} {\bibinfo {author} {\bibfnamefont {H.}~\bibnamefont
  {Fawzi}}, \bibinfo {author} {\bibfnamefont {J.}~\bibnamefont {Saunderson}}, \
  and\ \bibinfo {author} {\bibfnamefont {P.~A.}\ \bibnamefont {Parrilo}},\
  }\href {https://doi.org/10.1007/s10208-018-9385-0} {\bibfield  {journal}
  {\bibinfo  {journal} {Foundations of Computational Mathematics}\ }\textbf
  {\bibinfo {volume} {19}},\ \bibinfo {pages} {259} (\bibinfo {year} {2018})},\
  \bibinfo {note} {package cvxquad at
  \url{https://github.com/hfawzi/cvxquad}}\BibitemShut {NoStop}%
\bibitem [{\citenamefont {Grant}\ and\ \citenamefont {Boyd}(2014)}]{cvx}%
  \BibitemOpen
  \bibfield  {author} {\bibinfo {author} {\bibfnamefont {M.}~\bibnamefont
  {Grant}}\ and\ \bibinfo {author} {\bibfnamefont {S.}~\bibnamefont {Boyd}},\
  }\href@noop {} {\enquote {\bibinfo {title} {{CVX}: Matlab software for
  disciplined convex programming, version 2.1},}\ }\bibinfo {howpublished}
  {\url{http://cvxr.com/cvx}} (\bibinfo {year} {2014})\BibitemShut {NoStop}%
\bibitem [{\citenamefont {Grant}\ and\ \citenamefont {Boyd}(2008)}]{gb08}%
  \BibitemOpen
  \bibfield  {author} {\bibinfo {author} {\bibfnamefont {M.}~\bibnamefont
  {Grant}}\ and\ \bibinfo {author} {\bibfnamefont {S.}~\bibnamefont {Boyd}},\
  }in\ \href@noop {} {\emph {\bibinfo {booktitle} {Recent Advances in Learning
  and Control}}},\ \bibinfo {series and number} {Lecture Notes in Control and
  Information Sciences},\ \bibinfo {editor} {edited by\ \bibinfo {editor}
  {\bibfnamefont {V.}~\bibnamefont {Blondel}}, \bibinfo {editor} {\bibfnamefont
  {S.}~\bibnamefont {Boyd}}, \ and\ \bibinfo {editor} {\bibfnamefont
  {H.}~\bibnamefont {Kimura}}}\ (\bibinfo  {publisher} {Springer-Verlag
  Limited},\ \bibinfo {year} {2008})\ pp.\ \bibinfo {pages} {95--110},\
  \bibinfo {note} {\url{http://stanford.edu/~boyd/graph_dcp.html}}\BibitemShut
  {NoStop}%
\bibitem [{\citenamefont {Lin}\ \emph {et~al.}(2019)\citenamefont {Lin},
  \citenamefont {Upadhyaya},\ and\ \citenamefont
  {L{\"u}tkenhaus}}]{Lin:2019aa}%
  \BibitemOpen
  \bibfield  {author} {\bibinfo {author} {\bibfnamefont {J.}~\bibnamefont
  {Lin}}, \bibinfo {author} {\bibfnamefont {T.}~\bibnamefont {Upadhyaya}}, \
  and\ \bibinfo {author} {\bibfnamefont {N.}~\bibnamefont {L{\"u}tkenhaus}},\
  }\href@noop {} {\enquote {\bibinfo {title} {Asymptotic security analysis of
  discrete-modulated continuous-variable quantum key distribution},}\ }
  (\bibinfo {year} {2019}),\ \Eprint {http://arxiv.org/abs/arXiv:1905.10896}
  {arXiv:1905.10896} \BibitemShut {NoStop}%
\bibitem [{\citenamefont {H\"aseler}\ and\ \citenamefont
  {L\"utkenhaus}(2009)}]{Haseler09}%
  \BibitemOpen
  \bibfield  {author} {\bibinfo {author} {\bibfnamefont {H.}~\bibnamefont
  {H\"aseler}}\ and\ \bibinfo {author} {\bibfnamefont {N.}~\bibnamefont
  {L\"utkenhaus}},\ }\href {\doibase 10.1103/PhysRevA.80.042304} {\bibfield
  {journal} {\bibinfo  {journal} {Phys. Rev. A}\ }\textbf {\bibinfo {volume}
  {80}},\ \bibinfo {pages} {042304} (\bibinfo {year} {2009})}\BibitemShut
  {NoStop}%
\bibitem [{\citenamefont {Killoran}\ \emph {et~al.}(2012)\citenamefont
  {Killoran}, \citenamefont {Hosseini}, \citenamefont {Buchler}, \citenamefont
  {Lam},\ and\ \citenamefont {L\"utkenhaus}}]{Killoran12}%
  \BibitemOpen
  \bibfield  {author} {\bibinfo {author} {\bibfnamefont {N.}~\bibnamefont
  {Killoran}}, \bibinfo {author} {\bibfnamefont {M.}~\bibnamefont {Hosseini}},
  \bibinfo {author} {\bibfnamefont {B.~C.}\ \bibnamefont {Buchler}}, \bibinfo
  {author} {\bibfnamefont {P.~K.}\ \bibnamefont {Lam}}, \ and\ \bibinfo
  {author} {\bibfnamefont {N.}~\bibnamefont {L\"utkenhaus}},\ }\href {\doibase
  10.1103/PhysRevA.86.022331} {\bibfield  {journal} {\bibinfo  {journal} {Phys.
  Rev. A}\ }\textbf {\bibinfo {volume} {86}},\ \bibinfo {pages} {022331}
  (\bibinfo {year} {2012})}\BibitemShut {NoStop}%
\bibitem [{\citenamefont {Fowler}\ \emph {et~al.}(2012)\citenamefont {Fowler},
  \citenamefont {Mariantoni}, \citenamefont {Martinis},\ and\ \citenamefont
  {Cleland}}]{Fowler:2012aa}%
  \BibitemOpen
  \bibfield  {author} {\bibinfo {author} {\bibfnamefont {A.~G.}\ \bibnamefont
  {Fowler}}, \bibinfo {author} {\bibfnamefont {M.}~\bibnamefont {Mariantoni}},
  \bibinfo {author} {\bibfnamefont {J.~M.}\ \bibnamefont {Martinis}}, \ and\
  \bibinfo {author} {\bibfnamefont {A.~N.}\ \bibnamefont {Cleland}},\ }\href
  {\doibase 10.1103/PhysRevA.86.032324} {\bibfield  {journal} {\bibinfo
  {journal} {Phys. Rev. A}\ }\textbf {\bibinfo {volume} {86}},\ \bibinfo
  {pages} {032324} (\bibinfo {year} {2012})}\BibitemShut {NoStop}%
\end{thebibliography}%

\end{document}